\documentclass[a4paper]{IEEEtran}

\usepackage[dvips]{color}
\usepackage{graphicx}
\usepackage{amsmath}
\usepackage{mathrsfs}
\usepackage{amsfonts}
\usepackage{color}

\begin{document}
\title{Secrecy Energy Efficiency Optimization for Artificial Noise Aided Physical-Layer Security in OFDM-Based Cognitive Radio Networks}

\markboth{IEEE Transactions on Vehicular Technology (ACCEPTED TO APPEAR)}%
{Yuhan Jiang \emph{et al.}: Secrecy Energy Efficiency Optimization for Artificial Noise Aided Physical-Layer Security in OFDM-Based Cognitive Radio Networks}

\author{Yuhan~Jiang, Yulong~Zou,~\IEEEmembership{Senior Member,~IEEE}, Jian~Ouyang,~\IEEEmembership{Member,~IEEE}, and Jia~Zhu

\thanks{Copyright (c) 2015 IEEE. Personal use of this material is permitted. However, permission to use this material for any other purposes must be obtained from the IEEE by sending a request to pubs-permissions@ieee.org.}

\thanks{Manuscript received November 20, 2017; revised April 16, 2018, July 22, 2018, and September 20, 2018; accepted October 9, 2018. This work was partially supported by the National Natural Science Foundation of China under Grants 61522109, 61631020, 61671253, 91738201 and 61801234, the Natural Science Foundation of Jiangsu Province under Grants BK20150040, BK20160911 and BK20171446. \emph{(Corresponding author: Yulong Zou.)}}

\thanks{The authors are with the School of Telecommunications and Information Engineering, Nanjing University of Posts and Telecommunications, Nanjing 210003, China (e-mails: 15262769115@163.com; yulong.zou@njupt.edu.cn; ouyangjian@njupt.edu.cn; jiazhu@njupt.edu.cn).}

}

\maketitle

\begin{abstract}
In this paper, we investigate the power allocation of primary base station (PBS) and cognitive base station (CBS) across different orthogonal frequency division multiplexing (OFDM) subcarriers for energy-efficient secure downlink communication in OFDM-based cognitive radio networks (CRNs) with the existence of an eavesdropper having multiple antennas. For the sake of defending against eavesdropping, artificial noise is used to confuse the eavesdropper at the cost of extra power consumption. For the purpose of improving the energy efficiency (EE) of secure communications, we propose a secrecy energy efficiency maximization (SEEM) scheme by exploiting the instantaneous channel state information (ICSI) of the eavesdropper, called ICSI based SEEM (ICSI-SEEM) scheme with a given total transmit power budget for different OFDM subcarriers of both PBS and CBS while guaranteeing a certain secrecy rate (SR) for a cognitive user, where a primary user' SR is also taken into consideration for limiting the interference in CRNs at each subcarrier. As for the case when the eavesdropper's ICSI is unknown, we also propose an SEEM scheme through using the statistical CSI (SCSI) of the eavesdropper, namely SCSI based SEEM (SCSI-SEEM) scheme. Since the ICSI-SEEM and SCSI-SEEM problems are fractional and non-convex, we first transform them into equivalent subtractive problems, and then achieve approximate convex problems through employing the difference of two-convex functions approximation method. Finally, new two-tier power allocation algorithms are proposed to achieve $\varepsilon$-optimal solutions of our formulated ICSI-SEEM and SCSI-SEEM problems. Simulation results illustrate that the ICSI-SEEM has a better secrecy energy efficiency (SEE) performance than SCSI-SEEM, and moreover, the proposed ICSI-SEEM and SCSI-SEEM schemes outperform conventional SR maximization and EE maximization approaches in terms of their SEE performance.
\end{abstract}


\IEEEpeerreviewmaketitle

\section{Introduction}
In order to make full use of radio spectrum resources [1], extensive works have been devoted to investigating cognitive radio networks (CRNs), including cellular networks [2] and satellite networks [3]. In CRNs, the spectrum resources licensed to primary users (PUs) can be also allowed to cognitive users (CUs). Since the primary transmission dynamically changes with time between busy and idle states, the orthogonal frequency division multiplexing (OFDM) has been employed in CRNs by advantage of its flexibility in dynamic spectrum access [4]. However, even though CUs transmit over their detected spectrum holes in OFDM-based CRNs, the mutual interference between primary networks and CRNs still exists due to the occurrence of false alarm of a spectrum hole. Therefore, it is important to investigate power allocation for OFDM-based CRNs to control and limit such mutual interference below a tolerable level.

Besides, due to the broadcast nature of wireless networks, eavesdroppers (EDs) can overhear the confidential information transmitted over CRNs [5], which endangers the physical-layer security (PLS) of wireless communications seriously [6]. To defend against eavesdropping, many technologies have been utilized to ensure the secure transmission, including beamforming (BF) [7], artificial noise (AN) [8] and cooperative jamming [9], especially. Jamming can be used by the legitimate nodes to interfere with the EDs. Thus, it has a great potential in improving the transmission secrecy of wireless networks. For example, a cooperative jamming scheme has been presented for multi-antenna systems in [10]. Moreover, the authors also have optimized the power allocation between cooperative jammers to further improve the PLS. However, the improvement of secrecy performance is marginal when friendly jammers are near to legitimate receivers [11]. In such cases, the secrecy performance can be enhanced by employing BF technology [12], [13]. The secure BF design for multiuser multiple-input single-output (MISO) interference channel with an ED was investigated in [12]. The authors of [13] designed the secure BF to maximize the secrecy rate (SR) of secondary transmissions in an underlay MISO CRN, where broadcast channels are assumed to be overhead by massive EDs. The PLS of massive multiple-input multiple-output (MIMO) systems was enhanced in [14] by injecting the AN at the transmitter to interfere with the EDs at the cost of extra power consumption and exploiting the spatial degrees of freedom to guarantee the secure communication. In [15], AN was used in wiretap channels to improve the secrecy performance of three schemes, namely, the partially adaptive, fully adaptive, and ON-OFF schemes. The authors of [16] have studied the optimal power allocation for AN in wiretap channels with transmitter-side correlation to minimize the secrecy outage probability. In MISO wiretap channels with multiple antennas transmitter and single-antenna receiver and ED, AN was used for optimizing the secrecy performance in [17].

Also, since the energy resources are limited and most of them are not renewable, energy efficiency (EE) has been considered to be more and more important in CRNs, which is regarded as an efficient metric to balance the spectral efficiency (SE) and the power consumption [18]. In [19], the authors have studied a joint ergodic capacity maximization and average transmission power minimization problem for the secondary networks by employing spectrum sharing and spectrum sensing while satisfying PUs' quality-of-service (QoS). With the aid of cooperative jamming, EE was maximized through allocating power optimally under the constraints of secure transmission [20]. The authors of [21] investigated the physical layer power allocation and network layer delay in energy harvesting CRNs. For the aim of balancing the delay and EE, the delay power allocation was proposed and optimized. Considering the total power of CUs and interference of PUs, resource allocation problem in a multicarrier-based CRN was proposed to obtain the maximum CUs' EE in the condition of cooperative and uncooperative CUs [22].

Overall, the aforementioned research efforts [5]-[22] address either the case only concerned about SR or the case focused on EE. To this end, for the purpose of balancing the SR and EE better, the secrecy energy efficiency (SEE), has attracted considerable attention. To be specific, the SEE maximization (SEEM) problem was investigated in an underlay CRN which takes into account the transmit power constraint of cognitive base station (CBS) and SR of CU, at the same time, the QoS requirement of PU was also considered in [23]. The authors of [24] maximized the SEE of OFDM access (OFDMA) downlink network through allocating power, secrecy date rate and subcarrier resources subject to power consumption constraint and different QoS requirement. To take advantages of the cognitive radio and OFDM techniques, we study an SEEM problem for both instantaneous and statistical CSI of ED in a downlink OFDM-based CRN and propose an AN aided power allocation algorithm. The main contributions of this paper can be summarized as follows.
\begin{itemize}
\item We present a maximum ratio transmission (MRT) based confidential signal beamformer at CBS and propose an SEE optimization scheme for OFDM-based cognitive radio downlink transmissions. It is to maximize the SEE at the CBS by optimizing the power allocation between confidential and AN signals across different OFDM subcarriers with the total transmit power constraints for the primary base station (PBS) and CBS, while guaranteeing a required SR for the CU and PU.
\item We propose an SEEM scheme by exploiting the instantaneous CSI (ICSI) of the ED, namely ICSI based SEEM (ICSI-SEEM) scheme. However, the ICSI of ED may be unavailable in some cases. Therefore, we also propose an SEEM scheme by using the statistical CSI (SCSI) of the ED, called SCSI based SEEM (SCSI-SEEM) scheme.
\item Considering that our formulated ICSI-SEEM and SCSI-SEEM problems are fractional and non-convex, the original problems are converted into equivalent subtractive forms, and then they are transformed to convex problems by employing the difference of two-convex functions (D.C.) approximation method. Since there are no closed-form solutions for the proposed problems, new two-tier algorithms are proposed to achieve the corresponding $\varepsilon$-optimal power allocation solutions to our formulated problems.
\item Simulation results are given to prove the superiority of the proposed ICSI-SEEM and SCSI-SEEM schemes as well as the proposed MRT beamforming scheme with $\varepsilon$-optimal power allocation. Numerical results indicate that the proposed ICSI-SEEM and SCSI-SEEM schemes can balance the relationship between SR and EE better compared with the previous SR maximization (SRM) and EEM schemes. Moreover, the proposed schemes with $\varepsilon$-optimal power allocation algorithms obtain higher SEE than the other power allocation approaches.
\end{itemize}

The rest of the paper is organized as follows. In Section II, we describe the system model and introduce the performance metric used in this paper. Next, Section III formulates an SEE optimization problem with instantaneous CSI of ED for OFDM-based CRN systems and presents a two-tier algorithm to solve our formulated optimization problem. Then, in Section IV, we propose an SEEM problem with statistical CSI of ED and gives the corresponding solution, followed by Section V, where numerical simulation results are given to show the advantage of proposed SEEM schemes. Finally, a brief summary of our results are provided in Section V.

\emph{Notation:} Vectors or matrices are represented in bold letters. ${\rm E}( \cdot )$ represents the statistical expectation. ${\left(  \cdot  \right)^{\rm{H}}}$ denotes the conjugate transpose. The Euclidean norm of a vector is expressed as ${\left\|  \cdot  \right\|}$. ${\left[ x \right]^ + }$ is defined as $\max \left\{ {x,0} \right\}$. ${\rm{Tr}}\left( {\bf{A}} \right)$ is the trace of ${\bf{A}}$. ${{\bf{I}}_k}$ denotes the ${k\times k}$ identity matrix. ${{\mathbb{C}}^{N\times M}}$ is the space of all ${N\times M}$ matrices with complex entries. $\mathcal{CN}\left( 0,\sigma^{2} \right)$ represents a complex Gaussian random variable with zero mean and variance ${\sigma ^2}$.

\section{System Model and Performance Metric}
In this section, after presenting the system model used in this paper, we introduce the SEE as performance metric.
\subsection{System Model}

\begin{figure}
  \centering
  {\includegraphics[scale=0.6]{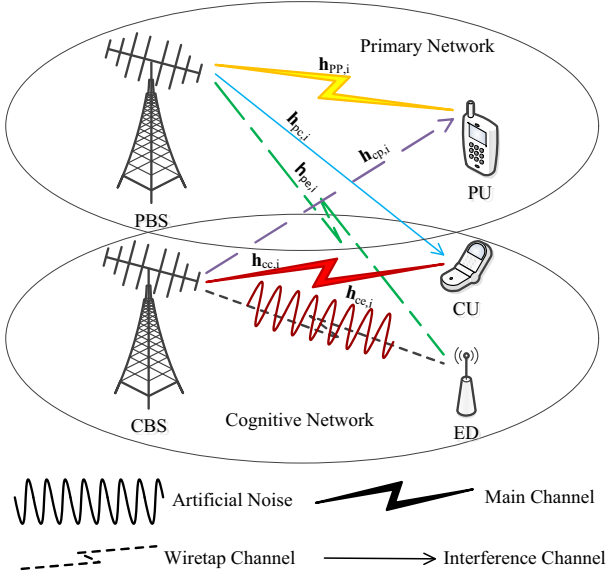}\\
  \caption{ System model for secure communication in OFDM-based CRN.}\label{Fig1}}
\end{figure}

We consider a downlink OFDM-based CRN having a CBS with $N_C$ antennas, a single-antenna CU and an ED with $N_E$ antennas coexists with a primary network (PN) having a PBS equipped with $N_P$ antennas and a single-antenna PU, as shown in Fig. 1. There are $I$ subcarriers in each OFDM symbol. On subcarrier $i \in \left\{ {1,...,I} \right\}$, CBS transmits confidential messages to CU with the same spectrum used by PN, where ED attempts to intercept the CBS-CU transmissions. To improve the PLS of cognitive transmissions, we adopt AN signals to confuse the ED.

At the ${i_{th}}$ subcarrier, the transmit signals of PBS and CBS can be respectively expressed by
\begin{equation}\tag{1}
{{\bf{x}}_{p,i}} = {{\bf{v}}_{p,i}}{p_i},
\end{equation}
\begin{equation}\tag{2}
{{\bf{x}}_{c,i}} = {{\bf{v}}_{s,i}}{s_i} + {{\bf{v}}_{z,i}}{z_i},
\end{equation}
where ${p_i}$ is the transmit signal of PBS and ${P_{p,i}} = {\rm E}\{ |{p_i}{|^2}\}$ is the transmit power of PBS at the ${i_{th}}$ subcarrier, ${{\bf{v}}_{p,i}}$ is the BF weight vector of the PBS' signal on subcarrier ${i}$, ${s_i}$ is the confidential signal, satisfying ${\rm{E}}\{ {\left| {{s_i}} \right|^2}\}  = {P_{s,i}}$ at the ${i_{th}}$ subcarrier, ${z_i}$ represents the AN signal with ${\rm{E}}\{ {\left| {z_i} \right|^2}\}  = {P_{z,i}}$ on subcarrier ${i}$, ${{\bf{v}}_{s,i}}$ and ${{\bf{v}}_{z,i}}$ are the BF weight vectors of the confidential and AN signals at the  ${i_{th}}$ subcarrier, respectively.

The received signals at PU, CU and ED on subcarrier ${i}$ can be respectively given by
\begin{equation}\tag{3}
{y_{p,i}} = {{\bf{h}}_{pp,i}}{{\bf{v}}_{p,i}}{p_i} + {{\bf{h}}_{cp,i}}{{\bf{v}}_{s,i}}{s_i} + {{\bf{h}}_{cp,i}}{{\bf{v}}_{z,i}}{z_i} + {n_{p,i}},
\end{equation}
\begin{equation}\tag{4}
{y_{c,i}} = {{\bf{h}}_{pc,i}}{{\bf{v}}_{p,i}}{p_i} + {{\bf{h}}_{cc,i}}{{\bf{v}}_{s,i}}{s_i} + {{\bf{h}}_{cc,i}}{{\bf{v}}_{z,i}}{z_i} + {n_{c,i}},
\end{equation}
\begin{equation}\tag{5}
{{\bf{y}}_{e,i}} = {{\bf{h}}_{pe,i}}{{\bf{v}}_{p,i}}{p_i} + {{\bf{h}}_{ce,i}}{{\bf{v}}_{s,i}}{s_i} + {{\bf{h}}_{ce,i}}{{\bf{v}}_{z,i}}{z_i} + {{\bf{n}}_{e,i}},
\end{equation}
where ${{\bf{h}}_{pp,i}}\in {{\mathbb{C}}^{1\times N_P}}$, ${{\bf{h}}_{pc,i}}\in {{\mathbb{C}}^{1\times N_P}}$ and ${{\bf{h}}_{pe,i}}\in {{\mathbb{C}}^{N_E\times N_P}}$ denote fading coefficients of the channel from PBS to PU, CU and ED at the ${i_{th}}$ subcarrier, respectively, ${{\bf{h}}_{cp,i}}\in {{\mathbb{C}}^{1\times N_C}}$, ${{\bf{h}}_{cc,i}}\in {{\mathbb{C}}^{1\times N_C}}$ and ${{\bf{h}}_{ce,i}}\in {{\mathbb{C}}^{N_E\times N_C}}$ are fading coefficients of the channel from CBS to PU, CU and ED, respectively at the ${i_{th}}$ subcarrier, ${{n}_{p,i}}\sim\mathcal{CN}\left( 0,{\sigma _{p,i}^2}\right)$, ${{n}_{c,i}}\sim\mathcal{CN}\left( 0,{\sigma _{c,i}^2} \right)$ and ${{\bf{n}}_{e,i}}\sim\mathcal{CN}\left( 0,{\sigma _{e,i}^2}{{\bf{I}}_{N_E}} \right)$ denote additive white Gaussian noises (AWGN) at PU, CU and ED on subcarrier ${i}$, respectively, with the same variance ${\sigma _{p,i}^2} = {\sigma _{c,i}^2} = {\sigma _{e,i}^2} = \Delta f{N_0}$, wherein $\Delta f$ and $N_0$ are the system bandwidth and single-sided noise spectral density, respectively.

From (3)-(4), the instantaneous signal-to-interference-plus-noise ratios (SINRs) at PU and CU on subcarrier ${i}$, respectively, can be written as
\begin{equation}\tag{6}
{\gamma _{p,i}} = \frac{{{{\left| {{{\bf{h}}_{pp,i}}{{\bf{v}}_{p,i}}} \right|}^2}{P_{p,i}}}}{{{{\left| {{{\bf{h}}_{cp,i}}{{\bf{v}}_{s,i}}} \right|}^2}{P_{s,i}} + {{\left| {{{\bf{h}}_{cp,i}}{{\bf{v}}_{z,i}}} \right|}^2}{P_{z,i}} + \sigma _{p,i}^2}},
\end{equation}
\begin{equation}\tag{7}
{\gamma _{c,i}} = \frac{{{{\left| {{{\bf{h}}_{cc,i}}{{\bf{v}}_{s,i}}} \right|}^2}{P_{s,i}}}}{{{{\left| {{{\bf{h}}_{pc,i}}{{\bf{v}}_{p,i}}} \right|}^2}{P_{p,i}} + {{\left| {{{\bf{h}}_{cc,i}}{{\bf{v}}_{z,i}}} \right|}^2}{P_{z,i}} + \sigma _{c,i}^2}},
\end{equation}
where the BF vector ${{\bf{v}}_{p,i}}$ and ${{\bf{v}}_{s,i}}$ are designed by MRT [25], i.e., ${{\bf{v}}_{p,i}} = \frac{{{\bf{h}}_{pp,i}^H}}{{\left\| {{{\bf{h}}_{pp,i}}} \right\|}}$ and ${{\bf{v}}_{s,i}} = \frac{{{\bf{h}}_{cc,i}^H}}{{\left\| {{{\bf{h}}_{cc,i}}} \right\|}}$. Meanwhile, for the purpose of guaranteeing that AN only degrades the channel condition of ED, we design ${{\bf{v}}_{z,i}}$ at the null space of ${{\bf{h}}_{cc,i}}$ and ${{\bf{h}}_{cp,i}}$, namely ${\bf{h}}_{cc,i}{{\bf{v}}_{z,i}} = 0$ and ${\bf{h}}_{cp,i}{{\bf{v}}_{z,i}} = 0$. Thus, the BF vector ${{\bf{v}}_{z,i}}$ is given by [26]
\begin{equation}\tag{8}
{{\bf{v}}_{z,i}} = \frac{{\Psi {\bf{h}}_{ce,i}^H}}{{\left\| {\Psi {\bf{h}}_{ce,i}^H} \right\|}}{\bf{w}},
\end{equation}
where $\Psi  = {{\bf{I}}_{{N_C}}} - \frac{{{\bf{h}}_i^H{{\bf{h}}_i}}}{{{{\left\| {{{\bf{h}}_i}} \right\|}^2}}}$, ${{\bf{h}}_i} = {\left[ {{{\bf{h}}_{cp,i}};{{\bf{h}}_{cc,i}}} \right]}$ and ${\bf{w}}$ is the AN vector ${{\bf{w}}}\sim\mathcal{CN}\left( 0,{\sigma _{e,i}^2}{{\bf{I}}_{N_E}} \right)$.

According to [27], the channel rates of PBS-ED and CBS-ED transmissions at the ${i_{th}}$ subcarrier can be respectively expressed as (9) and (10) at the top of the next page.

\subsection{Performance Metric}
The achievable SR of the CRN [28] is defined as
\begin{equation}\tag{11}
\begin{split}
{R_{\sec }}&\left( {{{\bf{P}}_p},{{\bf{P}}_s},{{\bf{P}}_z}} \right)\\ &= \sum\limits_{i = 1}^I {{{\left[ {R_{cc}}\left( {{{{P}}_{p,i}},{{P}_{s,i}},{{P}_{z,i}}} \right)- {R_{ce}}\left( {{{{P}}_{p,i}},{{P}_{s,i}},{{P}_{z,i}}} \right)\right]}^ + }},\\
\end{split}
\end{equation}
where ${{\bf{P}}_p} = \left[ {{P_{p,1}}{\kern 1pt} {\kern 1pt} {P_{p,2}}{\kern 1pt} {\kern 1pt}  \cdots {P_{p,I}}} \right]$, ${{\bf{P}}_s} = \left[ {{P_{s,1}}{\kern 1pt} {\kern 1pt} {P_{s,2}}{\kern 1pt} {\kern 1pt}  \cdots {P_{s,I}}} \right]$ and ${{\bf{P}}_z} = \left[ {{P_{z,1}}{\kern 1pt} {\kern 1pt} {P_{z,2}}{\kern 1pt} {\kern 1pt}  \cdots {P_{z,I}}} \right]$.

\begin{table*}[t]
\begin{equation}\tag{9}
{R_{pe}}\left( {{{{P}}_{p,i}},{{P}_{s,i}},{{P}_{z,i}}} \right) = {\log _2}\frac{{\left| {{{\bf{h}}_{pe,i}}{{\bf{v}}_{p,i}}{\bf{v}}_{p,i}^H{\bf{h}}_{pe,i}^H{P_{p,i}} + {{\bf{h}}_{ce,i}}{{\bf{v}}_{s,i}}{\bf{v}}_{s,i}^H{\bf{h}}_{ce,i}^H{P_{s,i}} + {{\bf{h}}_{ce,i}}{{\bf{v}}_{z,i}}{\bf{v}}_{z,i}^H{\bf{h}}_{ce,i}^H{P_{z,i}} + \sigma _{e,i}^2{{\bf{I}}_{{N_E}}}} \right|}}{{\left| {{{\bf{h}}_{ce,i}}{{\bf{v}}_{s,i}}{\bf{v}}_{s,i}^H{\bf{h}}_{ce,i}^H{P_{s,i}} + {{\bf{h}}_{ce,i}}{{\bf{v}}_{z,i}}{\bf{v}}_{z,i}^H{\bf{h}}_{ce,i}^H{P_{z,i}} + \sigma _{e,i}^2{{\bf{I}}_{{N_E}}}} \right|}},
\end{equation}
\end{table*}
\begin{table*}[t]
\begin{equation}\tag{10}
{R_{ce}}\left( {{{{P}}_{p,i}},{{P}_{s,i}},{{P}_{z,i}}} \right)= {\log _2}\frac{{\left| {{{\bf{h}}_{pe,i}}{{\bf{v}}_{p,i}}{\bf{v}}_{p,i}^H{\bf{h}}_{pe,i}^H{P_{p,i}} + {{\bf{h}}_{ce,i}}{{\bf{v}}_{s,i}}{\bf{v}}_{s,i}^H{\bf{h}}_{ce,i}^H{P_{s,i}} + {{\bf{h}}_{ce,i}}{{\bf{v}}_{z,i}}{\bf{v}}_{z,i}^H{\bf{h}}_{ce,i}^H{P_{z,i}} + \sigma _{e,i}^2{{\bf{I}}_{{N_E}}}} \right|}}{{\left| {{{\bf{h}}_{pe,i}}{{\bf{v}}_{p,i}}{\bf{v}}_{p,i}^H{\bf{h}}_{pe,i}^H{P_{p,i}} + {{\bf{h}}_{ce,i}}{{\bf{v}}_{z,i}}{\bf{v}}_{z,i}^H{\bf{h}}_{ce,i}^H{P_{z,i}} + \sigma _{e,i}^2{{\bf{I}}_{{N_E}}}} \right|}}.
\end{equation}
\hrule
\end{table*}

\begin{table*}[t]
\begin{equation}\renewcommand\theequation{15}
\begin{split}
\mathop {\max }\limits_{{{\bf{P}}_p},{{\bf{P}}_s},{{\bf{P}}_z}} {\kern 1pt} {\kern 1pt} {\kern 1pt} {\kern 1pt} {\kern 1pt} {\kern 1pt} {\kern 1pt} &{\eta _{{\rm{SEE}}}} = \frac{{\sum\limits_{i = 1}^I {\left[ {{{\log }_2}\left( {1 + \frac{{{e_i}{P_{s,i}}}}{{{b_i}{P_{p,i}} + \sigma _{c,i}^2}}} \right) - {{\log }_2}\frac{{\left| {{{\bf{c}}_i}{P_{p,i}} + {{\bf{f}}_i}{P_{s,i}} + {{\bf{g}}_i}{P_{z,i}} + \sigma _{e,i}^2{{\bf{I}}_{{N_E}}}} \right|}}{{\left| {{{\bf{c}}_i}{P_{p,i}} + {{\bf{g}}_i}{P_{z,i}} + \sigma _{e,i}^2{{\bf{I}}_{{N_E}}}} \right|}}} \right]} }}{{\sum\limits_{i = 1}^I {\left( {{P_{s,i}} + {P_{z,i}}} \right) + {P_b}} }}\\
{\kern 1pt} {\kern 1pt} {\kern 1pt} {\kern 1pt} {\kern 1pt} {\kern 1pt} {\kern 1pt} s.t.\quad  {\kern 1pt} {\kern 1pt} {\kern 1pt} {\kern 1pt} {\kern 1pt} {\kern 1pt} {\kern 1pt} {\kern 1pt} {\kern 1pt}& C1:{\log _2}\left( {1 + \frac{{{e_i}{P_{s,i}}}}{{{b_i}{P_{p,i}} + \sigma _{c,i}^2}}} \right) - {\log _2}\frac{{\left| {{{\bf{c}}_i}{P_{p,i}} + {{\bf{f}}_i}{P_{s,i}} + {{\bf{g}}_i}{P_{z,i}} + \sigma _{e,i}^2{{\bf{I}}_{{N_E}}}} \right|}}{{\left| {{{\bf{c}}_i}{P_{p,i}} + {{\bf{g}}_i}{P_{z,i}} + \sigma _{e,i}^2{{\bf{I}}_{{N_E}}}} \right|}} \ge R_{CU}^{\min },{\kern 1pt} {\kern 1pt} {\kern 1pt} {\kern 1pt} {\kern 1pt} {\kern 1pt} {\kern 1pt} \forall i,\\
{\kern 1pt} {\kern 1pt} {\kern 1pt} {\kern 1pt} {\kern 1pt} {\kern 1pt} {\kern 1pt} {\kern 1pt} {\kern 1pt} {\kern 1pt} {\kern 1pt} {\kern 1pt} {\kern 1pt} {\kern 1pt} {\kern 1pt} {\kern 1pt} {\kern 1pt} {\kern 1pt} {\kern 1pt} {\kern 1pt} {\kern 1pt} {\kern 1pt} {\kern 1pt} {\kern 1pt} {\kern 1pt} {\kern 1pt} {\kern 1pt} {\kern 1pt} {\kern 1pt} {\kern 1pt} &C2:{\log_2}\left( {1 + \frac{{{a_i}{P_{p,i}}}}{{{d_i}{P_{s,i}} + \sigma _{p,i}^2}}} \right) - {\log _2}\frac{{\left| {{{\bf{c}}_i}{P_{p,i}} + {{\bf{f}}_i}{P_{s,i}} + {{\bf{g}}_i}{P_{z,i}} + \sigma _{e,i}^2{{\bf{I}}_{{N_E}}}} \right|}}{{\left| {{{\bf{f}}_i}{P_{s,i}} + {{\bf{g}}_i}{P_{z,i}} + \sigma _{e,i}^2{{\bf{I}}_{{N_E}}}} \right|}} \ge R_{PU}^{\min },{\kern 1pt} {\kern 1pt} {\kern 1pt} {\kern 1pt} {\kern 1pt} {\kern 1pt} {\kern 1pt} \forall i,\\
{\kern 1pt} {\kern 1pt} {\kern 1pt} {\kern 1pt} {\kern 1pt} {\kern 1pt} {\kern 1pt} {\kern 1pt} {\kern 1pt} {\kern 1pt} {\kern 1pt} {\kern 1pt} {\kern 1pt} {\kern 1pt} {\kern 1pt} {\kern 1pt} {\kern 1pt} {\kern 1pt} {\kern 1pt} {\kern 1pt} {\kern 1pt} {\kern 1pt} {\kern 1pt} {\kern 1pt} {\kern 1pt} {\kern 1pt} {\kern 1pt} {\kern 1pt} {\kern 1pt} {\kern 1pt} {\kern 1pt} &C3:\sum\limits_{i = 1}^I {{P_{p,i}}}  \le P_{{\rm{PBS}}}^{{\rm{total}}},\\
{\kern 1pt} {\kern 1pt} {\kern 1pt} {\kern 1pt} {\kern 1pt} {\kern 1pt} {\kern 1pt} {\kern 1pt} {\kern 1pt} {\kern 1pt} {\kern 1pt} {\kern 1pt} {\kern 1pt} {\kern 1pt} {\kern 1pt} {\kern 1pt} {\kern 1pt} {\kern 1pt} {\kern 1pt} {\kern 1pt} {\kern 1pt} {\kern 1pt} {\kern 1pt} {\kern 1pt} {\kern 1pt} {\kern 1pt} {\kern 1pt} {\kern 1pt} {\kern 1pt} {\kern 1pt} {\kern 1pt}& C4:\sum\limits_{i = 1}^I {\left( {{P_{s,i}} + {P_{z,i}}} \right)}  \le P_{{\rm{CBS}}}^{{\rm{total}}},
\end{split}
\end{equation}
\hrule
\end{table*}

Besides, the total power consumption at the CBS can be modelled as [29]
\begin{equation}\tag{12}
{P_{\textrm{tot}}}({{{\bf{P}}_s}},{{{\bf{P}}_z}}) = \sum\limits_{i = 1}^I {\left( {{P_{s,i}} + {P_{z,i}}} \right)}  + {P_b},
\end{equation}
where ${P_b}$ is a constant circuit power consumed by the CBS.

Therefore, the SEE ${\eta _{\textrm{SEE}}}$ which measures the number of available secret bits transferred from the transmitter to receiver per unit energy and bandwidth of OFDM-based CRN systems can be expressed by [30]
\begin{equation}\tag{13}
{\eta _{\textrm{SEE}}} = \frac{{{R_{\textrm{sec} }}({{{\bf{P}}_p}},{{{\bf{P}}_s}},{{{\bf{P}}_z}})}}{{{P_{\textrm{tot}}}({{{\bf{P}}_s}},{{{\bf{P}}_z}})}}.
\end{equation}

\section{Secrecy Energy Efficiency Optimizations with Instantaneous CSI of ED}
In this section, we assume that the instantaneous CSI of ED is known, this CSI can be estimated by some technologies in some cases [31]-[33]. For example, we can estimate this CSI through local oscillator power leakage from the ED's radio frequency front-end [31]. Besides, if there exists an active ED in the wireless network, the CSI regarding the ED will be acquired [32]. Furthermore, due to the openness of wireless communications, some legal users may be captured by Trojan and then become EDs to wiretap the confidential transmissions. In this case, it is available to achieve the instantaneous CSI of the ED [33]. Therefore, we propose the eavesdropper's instantaneous CSI based SEEM (ICSI-SEEM) scheme. Then, due to the non-convexity of the proposed problem, we introduce the problem transformation. Finally, a two-tier power allocation algorithm is designed to obtain the $\varepsilon$-optimal SEE solution.

\subsection{Problem Formulation}
Our interest is to maximize SEE of the cognitive transmission under the SR constraints of CU and PU at each subcarrier and the total transmit power of PBS and CBS. Thus, the ICSI-SEEM can be formulated as

\begin{equation}\tag{14}
\begin{split}
&\mathop {\max }\limits_{{{\bf{P}}_p},{{\bf{P}}_s},{{\bf{P}}_z}}  \frac{{\sum\limits_{i = 1}^I {\left[ {R_{cc}}\left( {{{{P}}_{p,i}},{{P}_{s,i}},{{P}_{z,i}}} \right) - {R_{ce}}\left( {{{{P}}_{p,i}},{{P}_{s,i}},{{P}_{z,i}}} \right) \right]} }}{{\sum\limits_{i = 1}^I {\left( {{P_{s,i}} + {P_{z,i}}} \right) + {P_b}} }}\\
&s.t.C1\!:\!{R_{cc}}( {{{{P}}_{p,i}},{{P}_{s,i}},{{P}_{z,i}}}) \!-\! {R_{ce}}( {{{{P}}_{p,i}},{{P}_{s,i}},{{P}_{z,i}}} )\! \ge\! R_{CU}^{\min }, \forall i,\\
 &  {\kern 13pt}C2\!:\!{R_{pp}}( {{{{P}}_{p,i}},{{P}_{s,i}},{{P}_{z,i}}} )\! -\! {R_{pe}}( {{{{P}}_{p,i}},{{P}_{s,i}},{{P}_{z,i}}} ) \!\ge \!R_{PU}^{\min }, \forall i,\\
&{\kern 13pt}C3\!:\!\sum\limits_{i = 1}^I {{P_{p,i}}}  \le P_{{\rm{PBS}}}^{{\rm{total}}},\\
&{\kern 13pt}C4\!:\!\sum\limits_{i = 1}^I {\left( {{P_{s,i}} + {P_{z,i}}} \right)}  \le P_{{\rm{CBS}}}^{{\rm{total}}},
\end{split}
\end{equation}
where ${R_{pp}}({P_{p,i}},{P_{s,i}},{P_{z,i}}) = {\log _2}(1 + {\gamma _{p,i}})$ and ${R_{cc}}({P_{p,i}}, {P_{s,i}},$ ${P_{z,i}}) = {\log _2}(1 + {\gamma _{c,i}})$, $C1$ specifies the minimum SR requirement $R_{CU }^{\min }$ to ensure the security performance for CU at each subcarrier. For the sake of satisfying the SR requirement of PU, $C2$ gives a predefined threshold $R_{PU }^{\min }$ at the $i_{th}$ subcarrier to guarantee the PU' secure communications. Additionally, $C3$ and $C4$ are the transmit power constraints for PBS and CBS in the downlink OFDM-based CRN, where $P_{\textrm{PBS}}^{\textrm{total}}$ and  $P_{\textrm{CBS}}^{\textrm{total}}$ represent the maximum total transmit power of PBS and CBS, respectively.

Following [34]-[36], we can readily obtain the non-convexity of (14) due to its fractional form and logarithmic function, as shown from the objective function and constraint conditions in (14). It is challenging to solve a non-convex problem of (14). To this end, we introduce the following transformation.

\subsection{Problem Transformation}
Let ${a_i} = {\left| {{{\bf{h}}_{pp,i}}{{\bf{v}}_{p,i}}} \right|^2}$, ${b_i} = {\left| {{{\bf{h}}_{pc,i}}{{\bf{v}}_{p,i}}} \right|^2}$, ${{\bf{c}}_i} = {{\bf{h}}_{pe,i}}{{\bf{v}}_{p,i}}{\bf{v}}_{p,i}^H{\bf{h}}_{pe,i}^H$, ${d_i} = {\left| {{{\bf{h}}_{cp,i}}{{\bf{v}}_{s,i}}} \right|^2}$, ${e_i} = {\left| {{{\bf{h}}_{cc,i}}{{\bf{v}}_{s,i}}} \right|^2}$, ${{\bf{f}}_i} = {{\bf{h}}_{ce,i}}{{\bf{v}}_{s,i}}{\bf{v}}_{s,i}^H{\bf{h}}_{ce,i}^H$ and ${{\bf{g}}_i} = {{\bf{h}}_{ce,i}}{{\bf{v}}_{z,i}}{\bf{v}}_{z,i}^H{\bf{h}}_{ce,i}^H$, problem (14) can be formulated into (15) at the top of this page. Then, we are ready to introduce the following theorem.

\newtheorem{theorem}{Theorem}
\begin{theorem}
The optimal $\eta _{\textrm{SEE}}^ * $ for (15) can be acquired through the following optimization problem (16) if and only if $f(\eta _{\textrm{SEE}}^ * ) = 0$.
\end{theorem}
\begin{equation}\tag{16}
\begin{split}
&f( {{\eta _{\rm{SEE}}}})=\!\!\!\!\mathop {\max }\limits_{{{\bf{P}}_p},{{\bf{P}}_s},{{\bf{P}}_z}}\!\! \sum\limits_{i = 1}^I {\left[ {{f_1}\!\left( {{P_{p,i}},{P_{s,i}},{P_{z,i}}} \right)\! -\! {f_2}\!\left( {{P_{p,i}},{P_{s,i}},{P_{z,i}}} \right)} \right]} \\
&{\kern 1pt} \qquad \qquad - {\eta _{\textrm{SEE}}}\left[ {\sum\limits_{i = 1}^I {\left( {{P_{s,i}} + {P_{z,i}}} \right) + {P_b}} } \right]\\
&s.t.\quad\!\!\!\! C1\!:\!{f_1}\left( {{{P_{p,i}},{P_{s,i}},{P_{z,i}}} } \right) - {f_2}\left( {{{P_{p,i}},{P_{s,i}},{P_{z,i}}} } \right) \ge R_{CU }^{\textrm{min} },{\kern 1pt} {\kern 1pt} {\kern 1pt}  \forall i,\\
&\qquad\!\! C2\!:\!{g_1}\left( {{{P_{p,i}},{P_{s,i}},{P_{z,i}}} } \right) - {g_2}\left( {{{P_{p,i}},{P_{s,i}},{P_{z,i}}} } \right) \ge R_{PU }^{\textrm{min} },{\kern 1pt} {\kern 1pt} {\kern 1pt}  \forall i, \\
 &\qquad\!\!C3,\quad\!\!\! C4, \\
\end{split}
\end{equation}
where ${f_1}( {{P_{p,i}},{P_{s,i}},{P_{z,i}}} ) = {\log _2}( {{b_i}{P_{p,i}} + {e_i}{P_{s,i}} + \sigma _{c,i}^2} ) + $
${\log _2}\left| {{{\bf{c}}_i}{P_{p,i}} + {{\bf{g}}_i}{P_{z,i}} + \sigma _{e,i}^2{{\bf{I}}_{{N_E}}}} \right|$, ${f_2}( {{P_{p,i}},{P_{s,i}},{P_{z,i}}} ) = {\log _2}($ $ {{b_i}{P_{p,i}} + \sigma _{c,i}^2} ) + {\log _2}\left| {{{\bf{c}}_i}{P_{p,i}}} \right. + {{\bf{f}}_i}{P_{s,i}} + {{\bf{g}}_i}{P_{z,i}}$ $ + \left. {\sigma _{e,i}^2{{\bf{I}}_{{N_E}}}} \right|$, ${g_1}( {{P_{p,i}},{P_{s,i}},{P_{z,i}}} ) = {\log _2}( {{a_i}{P_{p,i}} + {d_i}{P_{s,i}} + \sigma _{p,i}^2} ) + {\log _2}|$ ${{{\bf{f}}_i}{P_{s,i}} + {{\bf{g}}_i}{P_{z,i}} + \sigma _{e,i}^2{{\bf{I}}_{{N_E}}}}|$ and ${g_2}( {{P_{p,i}},{P_{s,i}},{P_{z,i}}} ) = {\log _2}($ ${{d_i}{P_{s,i}} + \sigma _{p,i}^2}) + {\log _2}\left| {{{\bf{c}}_i}{P_{p,i}} + {{\bf{f}}_i}{P_{s,i}} + {{\bf{g}}_i}{P_{z,i}} + \sigma _{e,i}^2{{\bf{I}}_{{N_E}}}} \right|$.
\emph{Proof}: Please see Appendix A.

From Theorem 1, it is observed that the optimal solution of an optimization problem in fractional form can be solved by that in subtractive form. To this end, we will concentrate on solving the problem (16) in the rest of this paper.

\subsection{D.C. Programming}
Since the logarithmic functions ${f_1}({P_{p,i}},{P_{s,i}},{P_{z,i}})$, ${f_2}({P_{p,i}},$ ${P_{s,i}},{P_{z,i}})$, ${g_1}( {{P_{p,i}},{P_{s,i}},{P_{z,i}}} )$ and ${g_2}( {{P_{p,i}},{P_{s,i}},{P_{z,i}}})$ of (16) are concave, the functions ${f_1}-{f_2}$ and ${g_1}-{g_2}$ are D.C. functions, which become non-convex. For the purpose of solving the non-convex objective function, we apply the Taylor formula to approximate concave functions ${f_2}\left( {{P_{p,i}},{P_{s,i}},{P_{z,i}}} \right)$ and ${g_2}\left( {{P_{p,i}},{P_{s,i}},{P_{z,i}}} \right)$ into linear forms, which is the so-called D.C. approximation method [37]. The gradients of ${f_2}( {{P_{p,i}},{P_{s,i}},{P_{z,i}}} )$ and ${g_2}( {{P_{p,i}},{P_{s,i}},{P_{z,i}}} )$ are respectively given by
\begin{equation}\tag{17}
\begin{split}
d&{f_2}( {{P_{p,i}},{P_{s,i}},{P_{z,i}}} ) = \frac{{{b_i}}}{{( {{b_i}{P_{p,i}} + \sigma _{c,i}^2})\ln 2}}d{P_{p,i}}\\
& + \frac{{{\rm{Tr}}\left[ {{{\bf{c}}_i}{{( {{{\bf{c}}_i}{P_{p,i}} + {{\bf{f}}_i}{P_{s,i}} + {{\bf{g}}_i}{P_{z,i}} + \sigma _{e,i}^2{{\bf{I}}_{{N_E}}}} )}^{ - 1}}d{P_{p,i}}} \right]}}{{\ln 2}}\\
& + \frac{{{\rm{Tr}}\left[ {{{\bf{f}}_i}{{( {{{\bf{c}}_i}{P_{p,i}} + {{\bf{f}}_i}{P_{s,i}} + {{\bf{g}}_i}{P_{z,i}} + \sigma _{e,i}^2{{\bf{I}}_{{N_E}}}} )}^{ - 1}}d{P_{s,i}}} \right]}}{{\ln 2}}\\
& + \frac{{{\rm{Tr}}\left[ {{{\bf{g}}_i}{{( {{{\bf{c}}_i}{P_{p,i}} + {{\bf{f}}_i}{P_{s,i}} + {{\bf{g}}_i}{P_{z,i}} + \sigma _{e,i}^2{{\bf{I}}_{{N_E}}}} )}^{ - 1}}d{P_{z,i}}} \right]}}{{\ln 2}},
\end{split}
\end{equation}
and
\begin{equation}\tag{18}
\begin{split}
d&{g_2}( {{P_{p,i}},{P_{s,i}},{P_{z,i}}} ) = \frac{{{d_i}}}{{( {{d_i}{P_{s,i}} + \sigma _{p,i}^2} )\ln 2}}d{P_{s,i}} \\
&+ \frac{{{\rm{Tr}}\left[ {{{\bf{c}}_i}{{( {{{\bf{c}}_i}{P_{p,i}} + {{\bf{f}}_i}{P_{s,i}} + {{\bf{g}}_i}{P_{z,i}} + \sigma _{e,i}^2{{\bf{I}}_{{N_E}}}} )}^{ - 1}}d{P_{p,i}}} \right]}}{{\ln 2}}\\
& + \frac{{{\rm{Tr}}\left[ {{{\bf{f}}_i}{{( {{{\bf{c}}_i}{P_{p,i}} + {{\bf{f}}_i}{P_{s,i}} + {{\bf{g}}_i}{P_{z,i}} + \sigma _{e,i}^2{{\bf{I}}_{{N_E}}}} )}^{ - 1}}d{P_{s,i}}} \right]}}{{\ln 2}}\\
& + \frac{{{\rm{Tr}}\left[ {{{\bf{g}}_i}{{( {{{\bf{c}}_i}{P_{p,i}} + {{\bf{f}}_i}{P_{s,i}} + {{\bf{g}}_i}{P_{z,i}} + \sigma _{e,i}^2{{\bf{I}}_{{N_E}}}} )}^{ - 1}}d{P_{z,i}}} \right]}}{{\ln 2}},
\end{split}
\end{equation}
Then, according to the first-order Taylor series expansions of  ${f_2}( {{{{{P}}_{p,i}},{{{P}}_{s,i}},{{{P}}_{z,i}}}} )$ and ${g_2}( {{{{{P}}_{p,i}},{{{P}}_{s,i}},{{{P}}_{z,i}}}} )$, we have
\begin{equation}\tag{19}
\begin{split}
&{f_2}( {{P_{p,i}},{P_{s,i}},{P_{z,i}}}) \!\le\! {f_2}( {{{\bar P}_{p,i}},{{\bar P}_{s,i}},{{\bar P}_{z,i}}} ) \!+\! \frac{{{b_i}( {{P_{p,i}}\! -\! {{\bar P}_{p,i}}} )}}{{( {{b_i}{{\bar P}_{p,i}} \!+\! \sigma _{c,i}^2})\ln 2}} \\
&\!+\! \frac{{{\rm{Tr}}\left[ {{{\bf{c}}_i}{{( {{{\bf{c}}_i}{{\bar P}_{p,i}} \!+\! {{\bf{f}}_i}{{\bar P}_{s,i}}\! +\! {{\bf{g}}_i}{{\bar P}_{z,i}} \!+\! \sigma _{e,i}^2{{\bf{I}}_{{N_E}}}} )}^{ - 1}}\!( {{P_{p,i}}\! -\! {{\bar P}_{p,i}}})} \right]}}{{\ln 2}}\\
&\! +\! \frac{{{\rm{Tr}}\left[ {{{\bf{f}}_i}{{( {{{\bf{c}}_i}{{\bar P}_{p,i}}\! +\! {{\bf{f}}_i}{{\bar P}_{s,i}}\! +\! {{\bf{g}}_i}{{\bar P}_{z,i}}\! +\! \sigma _{e,i}^2{{\bf{I}}_{{N_E}}}} )}^{ - 1}}\!( {{P_{s,i}} \!- \!{{\bar P}_{s,i}}} )} \right]}}{{\ln 2}} \\
 &\!+\! \frac{{{\rm{Tr}}\left[ {{{\bf{g}}_i}{{( {{{\bf{c}}_i}{{\bar P}_{p,i}}\! +\! {{\bf{f}}_i}{{\bar P}_{s,i}}\! +\! {{\bf{g}}_i}{{\bar P}_{z,i}}\! +\! \sigma _{e,i}^2{{\bf{I}}_{{N_E}}}} )}^{ - 1}}\!( {{P_{z,i}}\! -\! {{\bar P}_{z,i}}})} \right]}}{{\ln 2}},
\end{split}
\end{equation}
and
\begin{equation*}
{g_2}( {{P_{p,i}},{P_{s,i}},{P_{z,i}}}) \!\le\! {g_2}( {{{\bar P}_{p,i}},{{\bar P}_{s,i}},{{\bar P}_{z,i}}}) \!+\! \frac{{{d_i}( {{P_{s,i}}\! -\! {{\bar P}_{s,i}}})}}{{\left( {{d_i}{{\bar P}_{s,i}} \!+\! \sigma _{p,i}^2} \right)\ln 2}}\\
\end{equation*}
\begin{equation}\tag{20}
\begin{split}
&\!+\! \frac{{{\rm{Tr}}\left[ {{{\bf{c}}_i}{{( {{{\bf{c}}_i}{{\bar P}_{p,i}} \!+\! {{\bf{f}}_i}{{\bar P}_{s,i}}\! +\! {{\bf{g}}_i}{{\bar P}_{z,i}}\! +\! \sigma _{e,i}^2{{\bf{I}}_{{N_E}}}} )}^{ - 1}}\!( {{P_{p,i}}\! -\! {{\bar P}_{p,i}}} )} \right]}}{{\ln 2}}\\
&\!+\! \frac{{{\rm{Tr}}\left[ {{{\bf{f}}_i}{{( {{{\bf{c}}_i}{{\bar P}_{p,i}}\! +\! {{\bf{f}}_i}{{\bar P}_{s,i}}\! + \!{{\bf{g}}_i}{{\bar P}_{z,i}}\! +\! \sigma _{e,i}^2{{\bf{I}}_{{N_E}}}} )}^{ - 1}}\!( {{P_{s,i}}\! -\! {{\bar P}_{s,i}}} )} \right]}}{{\ln 2}}\\
 &\! +\! \frac{{{\rm{Tr}}\left[ {{{\bf{g}}_i}{{( {{{\bf{c}}_i}{{\bar P}_{p,i}}\! +\! {{\bf{f}}_i}{{\bar P}_{s,i}}\! +\! {{\bf{g}}_i}{{\bar P}_{z,i}}\! +\! \sigma _{e,i}^2{{\bf{I}}_{{N_E}}}} )}^{ - 1}}\!( {{P_{z,i}}\! -\! {{\bar P}_{z,i}}} )} \right]}}{{\ln 2}},
\end{split}
\end{equation}
where $\left({{{{{\bar P}}}_{p,i}},{{{{\bar P}}}_{s,i}},{{{{\bar P}}}_{z,i}}} \right)$ is a feasible solution of ${f_2}({{{P}}_{p,i}},{{{P}}_{s,i}},$ ${{{P}}_{z,i}})$ and ${g_2}\left({{{{P}}_{p,i}},{{{P}}_{s,i}},{{{P}}_{z,i}}}\right)$. By substituting (19) and (20) into the problem (16), and denoting ${{{\bf{\bar \Omega }}_i}} = {{\bf{c}}_i}{{\bar P}_{p,i}} + {{\bf{f}}_i}{{\bar P}_{s,i}} + {{\bf{g}}_i}{{\bar P}_{z,i}} + \sigma _{e,i}^2{{\bf{I}}_{{N_E}}}$, we can reformulate (16) as
\begin{equation}\tag{21}
\begin{split}
&\mathop {\max }\limits_{{{\bf{P}}_p},{{\bf{P}}_s},{{\bf{P}}_z}}\sum\limits_{i = 1}^I{\left\{ {f_1}\left( {{P_{p,i}},{P_{s,i}},{P_{z,i}}} \right) \right.}\! -\!{f_2}( {{{\bar P}_{p,i}},{{\bar P}_{s,i}},{{\bar P}_{z,i}}})\!\\
& -\! \frac{{{b_i}({P_{p,i}} - {{\bar P}_{p,i}})}}{{({b_i}{{\bar P}_{p,i}} + \sigma _{c,i}^2)\ln 2}}\! -\! \frac{{ {{\rm{Tr}}\left[ {{{\bf{c}}_i}{{( {\bf{\bar \Omega }}_i)}^{ - 1}}( {{P_{p,i}} - {{\bar P}_{p,i}}})} \right]} }}{{\ln 2}}\\
& -\! \frac{{ {{\rm{Tr}}\left[ {{{\bf{f}}_i}{{({\bf{\bar \Omega }}_i)}^{ - 1}}( {{P_{s,i}}\! -\! {{\bar P}_{s,i}}} )} \right]} }}{{\ln 2}} - \!\!\left. {\frac{{ {{\rm{Tr}}\left[ {{{\bf{g}}_i}{{( {\bf{\bar \Omega }}_i)}^{ - 1}}( {{P_{z,i}}\! -\! {{\bar P}_{z,i}}} )} \right]} }}{{\ln 2}}} \!\right\} \\
&- {\eta _{{\rm{SEE}}}}\left[ {\sum\limits_{i = 1}^I {\left( {{P_{s,i}} + {P_{z,i}}} \right) + {P_b}} } \right]\\
& s.t.{f_1}\left( {{P_{p,i}},{P_{s,i}},{P_{z,i}}} \right)\! -\! {f_2}\left( {{{\bar P}_{p,i}},{{\bar P}_{s,i}},{{\bar P}_{z,i}}} \right)\!-\! \frac{{{b_i}( {{P_{p,i}}\! -\! {{\bar P}_{p,i}}} )}}{{( {{b_i}{{\bar P}_{p,i}}\!+\! \sigma _{c,i}^2} )\ln 2}}  \\
&\qquad- \frac{{{\rm{Tr}}\left[ {{{\bf{c}}_i}{{ {{\bf{\bar\Omega }}_i}}^{ - 1}}( {{P_{p,i}}\! -\! {{\bar P}_{p,i}}} )} \right]}}{{\ln 2}} -  \frac{{{\rm{Tr}}\!\left[ {{{\bf{f}}_i}{{ {{\bf{\bar\Omega }}_i} }^{ - 1}}( {{P_{s,i}}\! -\! {{\bar P}_{s,i}}} )} \right]}}{{\ln 2}}\\
& \qquad-\! \frac{{{\rm{Tr}}\!\left[ {{{\bf{g}}_i}{{ {{\bf{\bar\Omega }}_i}}^{ - 1}}({{P_{z,i}} \!-\! {{\bar P}_{z,i}}} )} \right]}}{{\ln 2}} \!\ge\! R_{CU}^{\min },{\kern 1pt} \forall i,\\
&\qquad\!\! {g_1}( {{P_{p,i}},{P_{s,i}},{P_{z,i}}} )\! -\! {g_2}( {{{\bar P}_{p,i}},{{\bar P}_{s,i}},{{\bar P}_{z,i}}} )\!- \!\frac{{{d_i}( {{P_{s,i}}\! -\! {{\bar P}_{s,i}}} )}}{{( {{d_i}{{\bar P}_{s,i}}\! +\! \sigma _{p,i}^2} )\ln 2}} \\
& \qquad- \frac{{{\rm{Tr}}\left[ {{{\bf{c}}_i}{{{ {{\bf{\bar\Omega }}_i}}}^{ - 1}}( {{P_{p,i}}\! -\! {{\bar P}_{p,i}}} )} \right]}}{{\ln 2}}- \frac{{{\rm{Tr}}\!\left[ {{{\bf{f}}_i}{{{ {{\bf{\bar\Omega }}_i}}}^{ - 1}}( {{P_{s,i}}\! -\! {{\bar P}_{s,i}}} )} \right]}}{{\ln 2}} \\
&\qquad-\! \frac{{{\rm{Tr}}\!\left[ {{{\bf{g}}_i}{{{ {{\bf{\bar\Omega }}_i}}}^{ - 1}}( {{P_{z,i}}\! -\! {{\bar P}_{z,i}}} )} \right]}}{{\ln 2}}\! \ge\! R_{PU}^{\min },{\kern 1pt}\forall i,\\
&\qquad\quad\!\!\!\!C3,\quad\!\!\! C4.
\end{split}
\end{equation}
Following [38] and [39], it is obvious that the problem (21) is convex, which results from the convexity of the objective function as well as that of the constraints $C1$, $C2$, $C3$ and $C4$. Therefore, it is simple and straightforward to obtain the optimal solution to (21) by using existing convex software tools, e.g., CVX [40].

Based on (21), we propose the following iterative procedure, which converges to the optimal solutions of problem (16).
\begin{equation*}
\begin{split}
&({\bf{\bar P}}_p^{n + 1},{\bf{\bar P}}_s^{n + 1},{\bf{\bar P}}_z^{n + 1}) = \\
&=\arg\mathop {\max }\limits_{{{\bf{P}}_p},{{\bf{P}}_s},{{\bf{P}}_z}} \sum\limits_{i = 1}^I {\left\{ {f_1}\left( {{P_{p,i}},{P_{s,i}},{P_{z,i}}} \right) \right.}\! -\!{f_2}( {{{\bar P}_{p,i}^n},{{\bar P}_{s,i}^n},{{\bar P}_{z,i}^n}})\!\\
& -\! \frac{{{b_i}({P_{p,i}} - {{\bar P}_{p,i}^n})}}{{({b_i}{{\bar P}_{p,i}^n} + \sigma _{c,i}^2)\ln 2}}\! -\! \frac{{ {{\rm{Tr}}\left[ {{{\bf{c}}_i}{{( {\bf{\bar \Omega }^n}_i)}^{ - 1}}( {{P_{p,i}}\! -\! {{\bar P}_{p,i}^n}})} \right]} }}{{\ln 2}}\\
\end{split}
\end{equation*}
\begin{equation}\tag{22}
\begin{split}
& -\! \frac{{ {{\rm{Tr}}\left[ {{{\bf{f}}_i}{{({\bf{\bar \Omega }^n}_i)}^{ - 1}}( {{P_{s,i}}\! -\! {{\bar P}_{s,i}^n}} )} \right]} }}{{\ln 2}} - \!\!\left. {\frac{{ {{\rm{Tr}}\left[ {{{\bf{g}}_i}{{( {\bf{\bar \Omega }^n}_i)}^{ - 1}}( {{P_{z,i}} - {{\bar P}_{z,i}^n}} )} \right]} }}{{\ln 2}}} \!\right\} \\
&- {\eta _{{\rm{SEE}}}}\left[ {\sum\limits_{i = 1}^I {\left( {{P_{s,i}} + {P_{z,i}}} \right) + {P_b}} } \right]\\
& s.t.{f_1}( {{{{P}}_{p,i}},{{{P}}_{s,i}},{{{P}}_{z,i}}} )\! -\!{f_2}({{\bar P}}_{p,i}^n,{{\bar P}}_{s,i}^n,{{\bar P}}_{z,i}^n)\! -\! \frac{{{b_i}\left( {{P_{p,i}}\! -\! \bar P_{p,i}^n} \right)}}{{\left( {{b_i}\bar P_{p,i}^n\! +\! \sigma _{c,i}^2} \right)\ln 2}}\\
& -\frac{{{\rm{Tr}}\left[ {{{\bf{c}}_i}{{\left( {\bf{\bar \Omega }}_i^n \right)}^{ - 1}}\!\left( {{P_{p,i}}\! -\! \bar P_{p,i}^n} \right)} \right]}}{{\ln 2}} \!- \!\frac{{{\rm{Tr}}\left[ {{{\bf{f}}_i}{{\left({\bf{\bar \Omega }}_i^n \right)}^{ - 1}}\!\left( {{P_{s,i}}\! -\! \bar P_{s,i}^n} \right)} \right]}}{{\ln 2}}\\
& - {\frac{{{\rm{Tr}}\left[ {{{\bf{g}}_i}{{\left( {\bf{\bar \Omega }}_i^n\right)}^{ - 1}}\!\left( {{P_{z,i}}\! -\! \bar P_{z,i}^n} \right)} \right]}}{{\ln 2}}} \ge R_{CU}^{\min },{\kern 1pt}\forall i,\\
&\quad{g_1}( {{{{P}}_{p,i}},{{{P}}_{s,i}},{{{P}}_{z,i}}} )\! -\!{g_2}({{\bar P}}_{p,i}^n,{{\bar P}}_{s,i}^n,{{\bar P}}_{z,i}^n  )\!-\!\frac{{{d_i}\left( {{P_{s,i}} \!-\! \bar P_{s,i}^n} \right)}}{{\left( {{d_i}\bar P_{s,i}^n \!+\! \sigma _{p,i}^2} \right)\ln 2}} \\
& - \frac{{{\rm{Tr}}\left[ {{{\bf{c}}_i}{{\left({\bf{\bar \Omega }}_i^n \right)}^{ - 1}}\!\left( {{P_{p,i}}\! -\! \bar P_{p,i}^n} \right)} \right]}}{{\ln 2}} \!-\! \frac{{{\rm{Tr}}\left[ {{{\bf{f}}_i}{{\left({\bf{\bar \Omega }}_i^n \right)}^{ - 1}}\!\left( {{P_{s,i}}\! -\! \bar P_{s,i}^n} \right)} \right]}}{{\ln 2}}\\
&-{\frac{{{\rm{Tr}}\left[ {{{\bf{g}}_i}{{\left( {\bf{\bar \Omega }}_i^n \right)}^{ - 1}}\!\left( {{P_{z,i}}\! -\! \bar P_{z,i}^n} \right)} \right]}}{{\ln 2}}} \ge R_{PU}^{\min },{\kern 1pt}\forall i,\\
&\quad C3,\quad\!\!\! C4,
\end{split}
\end{equation}
where ${\bf{\bar \Omega }}_i^n={{{\bf{c}}_i}\bar P_{p,i}^n\!+\! {{\bf{f}}_i}\bar P_{s,i}^n \!+\! {{\bf{g}}_i}\bar P_{z,i}^n \!+\! \sigma _{e,i}^2{{\bf{I}}_{{N_E}}}}$, $( {\bf{\bar P}}_p^n,{\bf{\bar P}}_s^n,{\bf{\bar P}}_z^n )$ and $( {\bf{\bar P}}_p^{n + 1},{\bf{\bar P}}_s^{n + 1},{\bf{\bar P}}_z^{n + 1} )$ the optimal solutions in (22) at iterations $n$ and $n+1$, respectively.

\begin{proof}
Please see Appendix B for the proof of convergence.
\end{proof}

\subsection{Two-tier Iterative Algorithm for ICSI-SEEM}
In this section, we propose a two-tier iterative power allocation algorithm to obtain an $\varepsilon$-optimal power allocation solution to our formulated ICSI-SEEM problem. The proposed algorithm is summarized in Table I. First of all, we initialize the maximum SEE ${\eta _{\textrm{SEE}}^{m}} = 0$ and iteration index $m = 0$, $n = 0$. Based on the given maximum SEE ${\eta _{\textrm{SEE}}^{m}}$ at the outer tier, the D.C. approximation method is applied to solve problem (16) for obtaining the $\varepsilon$-optimal solution $({\bf{P}}_p^{n},{\bf{P}}_s^{n},{\bf{P}}_z^{n})$ at the inner tier. The $\varepsilon$-optimal solution  $({\bf{P}}_p^{n},{\bf{P}}_s^{n},{\bf{P}}_z^{n})$ will be used to update the value of $f( {{\eta _{\textrm{SEE}}}} )$ for the next outer tier. Meanwhile, ${\eta_{\textrm{SEE}}}$ is found to satisfy $f( {{\eta _{\textrm{SEE}}}}) = 0$ by using the Dinkelbach's method [41] at this tier. When all the updated data nearly keeps unchanged or the number of iterations approaches to the maximization, the iteration stops; otherwise, another round of iteration starts.

\vspace{0.05in}
\begin{table}[!h]
\caption{Two-tier Iterative $\varepsilon$-optimal Power Allocation Algorithm for ICSI-SEEM Scheme}
\begin{center}
\begin{tabular}{l}
\hline
Algorithm 1: Two-tier Iterative $\varepsilon$-optimal Power Allocation Algorithm.\\
\hline
\textbf{Function}
\emph{Outer{\_} Iteration}\\
\small
Step 1: Initialize the maximum number of iterations ${m_{\textrm{max} }}$, ${n_{\textrm{max} }}$\\
~~~~~~~~and the maximum tolerance $\varepsilon$.\\
\small
Step 2: Set maximum SEE $\eta _{\textrm{SEE}}^0 = 0$ and iteration index
 $m=0$.\\
\small
Step 3: Call
\textbf{Function}
\emph{Inner{\_} Iteration} with $\eta _{\textrm{SEE}}^m$ to obtain the\\
{\kern 25pt}$\varepsilon$-optimal solution $( {{\bf{P}}_{p}^n,{\bf{P}}_{s}^n,{\bf{P}}_{z}^n} )$.\\
\small
Step 4: Update $\eta _{{\rm{SEE}}}^{m + 1}\!$ =\\
$\quad \frac{{\sum\limits_{i = 1}^I\!\! {\left[ {{{\log }_2}\left( {1 + \frac{{{e_i}P_{s,i}^n}}{{{b_i}P_{p,i}^n + \sigma _{c,i}^2}}} \right)\! - {{\log}_2}\left( {\frac{{\left| {{{\bf{c}}_i}P_{p,i}^n + {{\bf{f}}_i}P_{s,i}^n + {{\bf{g}}_i}P_{z,i}^n + \sigma _{e,i}^2{{\bf{I}}_{{N_E}}}} \right|}}{{\left| {{{\bf{c}}_i}P_{p,i}^n + {{\bf{g}}_i}P_{z,i}^n + \sigma _{e,i}^2{{\bf{I}}_{{N_E}}}} \right|}}} \right)} \right]} }}{{\sum\limits_{i = 1}^I {\left( {P_{s,i}^n + P_{z,i}^n} \right)}  + {P_b}}}$.\\
\small
Step 5: Set $m=m+1$.\\
\small
Step 6: \textbf{if} $\left| {\eta _{\textrm{SEE}}^m - \eta _{\textrm{SEE}}^{m - 1}} \right| \ge \varepsilon $ or $m \le {m_{\textrm{max} }}$ \\
\small
Step 7: \textbf{goto} Step 3.\\
\small
Step 8: \textbf{end if}\\
\small
Step 9: \textbf{return} ${\bf{P}}_p^{n}$, ${\bf{P}}_s^{n}$, ${\bf{P}}_z^{n}$.\\
\small
Step 10: Obtain the $\varepsilon$-optimal solution ${\bf{P}}_{p}^ *  = {\bf{P}}_{p}^n$, ${\bf{P}}_{s}^ *  = {\bf{P}}_{s}^n$\\
{\kern 30pt} and ${\bf{P}}_{z}^ *  = {\bf{P}}_{z}^n$ for problem (15).\\
\small
\textbf{end} \\
\small
\textbf{Function}
\emph{Inner{\_} Iteration}  $(\eta _{\textrm{SEE}})$\\
\small
Step 11: Initialize $({\bf{P}} _{p}^0, {\bf{P}} _{s}^0, {\bf{P}} _{z}^0) = (0,0,0)$ and ${f^0} = 0$.\\
\small
Step 12: Set $n = 0$.\\
\small
Step 13: Find the $\varepsilon$-optimal solution $\left( {{{\bf{P}}_{p}^{n+1}},{{\bf{P}}_{s}^{n+1}},{{\bf{P}}_{z}^{n+1}}} \right)$ of (22)\\
{\kern 30pt} for given $( {{{\bf{P}}_{p}^{n}},{{\bf{P}}_{s}^{n}},{{\bf{P}}_{z}^{n}}}  )$ and ${\eta _{\textrm{SEE}}^m}$ by using CVX.\\
\small
Step 14: Compute\\
${f^{n + 1}} = \sum\limits_{i = 1}^I {\left[ {{f_1}\left( {P_{p,i}^{n + 1},P_{s,i}^{n + 1},P_{z,i}^{n + 1}} \right) - {f_2}\left( {P_{p,i}^{n + 1},P_{s,i}^{n + 1},P_{z,i}^{n + 1}} \right)} \right]}  $\\
$\qquad\qquad - {\eta _{\textrm{SEE}}^m}\left[ {\sum\limits_{i = 1}^I {\left( { P_{s,i}^{n + 1} +  P_{z,i}^{n + 1}} \right) + {P_b}} } \right].$\\
\small
Step 15: Set $n=n+1$.\\
\small
Step 16: \textbf{if} $\left| {{f^n} - {f^{n - 1}}} \right| \ge \varepsilon $ or $n \le {n_{\textrm{max} }}$  \\
\small
Step 17: \textbf{goto} Step 13.\\
\small
Step 18: \textbf{end if}\\
\small
Step 19: \textbf{return} ${{\bf{P}}_{p}^{n}}$, ${{\bf{P}}_{s}^{n}}$,     ${{\bf{P}}_{z}^{n}}$.\\
\small
\textbf{end}\\
\hline
\end{tabular}
\end{center}
\end{table}

The computational complexity of the proposed scheme depends on the number of iterations, variable size and the number of constraints at the outer and inner tiers. Based on the given tolerance $\varepsilon $, we can give the iterations as $O\left( {\log \left( {\eta _{\textrm{SEE}}^{up}/\varepsilon } \right)\log \left( {g_{\textrm{SEE}}^{up}/\varepsilon } \right)} \right)$, where $\eta _{\textrm{SEE}}^{up} = ( {\frac{{\max ({e_i})P_{\textrm{CBS}}^{\textrm{total}}}}{{\Delta f{N_0}\ln 2}}} )/{P_b}$ and $g_{\textrm{SEE}}^{up} = \frac{{\max ({e_i})P_{\textrm{CBS}}^{\textrm{total}}}}{{\Delta f{N_0}\ln 2}}$. Given $3I$ scalar variables in problem (22), so we need at most $O((3I)^{3.5}\log (1/\varepsilon ))$ calculations at each inner iteration [42]. Finally, the overall computational complexity of the proposed scheme can be roughly written as
\begin{equation}\tag{23}
O\left( {\log \left( {\frac{1}{\varepsilon }} \right)\log \left( {\frac{{\eta _{{\rm{SEE}}}^{up}}}{\varepsilon }} \right)\log \left( {\frac{{g_{\rm{SEE}}^{up}}}{\varepsilon }} \right){{\left( {3I} \right)}^{3.5}}} \right).
\end{equation}

\section{Secrecy Energy Efficiency Optimizations with Statistical CSI of ED}
For the reason that the instantaneous CSI of ED may be unavailable in some cases, we propose an SEEM scheme through using the statistical CSI of the ED [43], [44], namely the eavesdropper's statistical CSI based SEEM (SCSI-SEEM) scheme in this section. Then, we give the solution of our formulated SCSI-SEEM problem. Finally, a two-tier iterative $\varepsilon$-optimal power allocation algorithm is presented for SCSI-SEEM scheme.
\subsection{SCSI-SEEM Problem Formulation}
We formulate the SCSI-SEEM problem in OFDM-based CRNs as
\begin{equation*}
\begin{split}
&\mathop {\max }\limits_{{{\bf{P}}_p},{{\bf{P}}_s},{{\bf{P}}_z}} \frac{{\sum\limits_{i = 1}^I {\left\{ {{R_{cc}}({P_{p,i}},{P_{s,i}},{P_{z,i}}) - {\rm{E[}}{R_{ce}}({P_{p,i}},{P_{s,i}},{P_{z,i}})]} \right\}} }}{{\sum\limits_{i = 1}^I {\left( {{P_{s,i}} + {P_{z,i}}} \right) + {P_b}} }}\\
& s.t. C1\!:\!{R_{cc}}({P_{p,i}},\!{P_{s,i}},\!{P_{z,i}})\! -\! {\rm{E[}}{R_{ce}}({P_{p,i}},\!{P_{s,i}},\!{P_{z,i}})] \!\ge\! R_{CU}^{\min },\forall i,\\
& \quad {\kern 3pt} C2\!:\!{R_{pp}}({P_{p,i}},\!{P_{s,i}},\!{P_{z,i}})\! -\! {\rm{E[}}{R_{pe}}({P_{p,i}},\!{P_{s,i}},\!{P_{z,i}})] \!\ge\! R_{PU}^{\min }, \forall i,\\
\end{split}
\end{equation*}
\begin{equation}\tag{24}
\begin{split}
& \quad {\kern 3pt} C3\!:\!\sum\limits_{i = 1}^I {{P_{p,i}}}  \le P_{{\rm{PBS}}}^{{\rm{total}}},\qquad\qquad\qquad\quad\\
& \quad {\kern 3pt} C4\!:\!\sum\limits_{i = 1}^I {\left( {{P_{s,i}} + {P_{z,i}}} \right) \le P_{{\rm{CBS}}}^{{\rm{total}}}.}\qquad\qquad\qquad\quad
\end{split}
\end{equation}
After some operations, problem (24) can be rewritten as (25) at top of the next page.

\subsection{SCSI-SEEM Solution}
According to Theorem 1, we can achieve the optimal solution $\varphi _{{\rm{SEE}}}^ * $ of (25) through problem (26) if and only if $h( {\varphi _{{\rm{SEE}}}^ *} ) = 0$.
\begin{equation}\tag{26}
\begin{split}
&h( {{\varphi_{\rm{SEE}}}})=\!\!\!\!\mathop {\max }\limits_{{{\bf{P}}_p},{{\bf{P}}_s},{{\bf{P}}_z}}\!\! \sum\limits_{i = 1}^I {\left[ {{h_1}\!\left( {{P_{p,i}},{P_{s,i}},{P_{z,i}}} \right)\! -\! {h_2}\!\left( {{P_{p,i}},{P_{s,i}},{P_{z,i}}} \right)} \right]} \\
&{\kern 1pt} \qquad \qquad - {\varphi _{\textrm{SEE}}}\left[ {\sum\limits_{i = 1}^I {\left( {{P_{s,i}} + {P_{z,i}}} \right) + {P_b}} } \right]\\
&s.t.\quad\!\!\!\! C1\!:\!{h_1}\left( {{{P_{p,i}},{P_{s,i}},{P_{z,i}}} } \right) - {h_2}\left( {{{P_{p,i}},{P_{s,i}},{P_{z,i}}} } \right) \ge R_{CU }^{\textrm{min} },{\kern 1pt} {\kern 1pt} {\kern 1pt}  \forall i,\\
&\qquad\!\! C2\!:\!{r_1}\left( {{{P_{p,i}},{P_{s,i}},{P_{z,i}}} } \right) - {r_2}\left( {{{P_{p,i}},{P_{s,i}},{P_{z,i}}} } \right) \ge R_{PU }^{\textrm{min} },{\kern 1pt} {\kern 1pt} {\kern 1pt}  \forall i, \\
 &\qquad\!\!C3,\quad\!\!\! C4, \\
\end{split}
\end{equation}
where ${h_1}\left( {{P_{p,i}},{P_{s,i}},{P_{z,i}}} \right) = {\log _2}\left( {{b_i}{P_{p,i}} + {e_i}{P_{s,i}} + \sigma _{c,i}^2} \right) + $
$\rm{E}[{\log _2}\left| {{{\bf{c}}_i}{P_{p,i}} + {{\bf{g}}_i}{P_{z,i}} + \sigma _{e,i}^2{{\bf{I}}_{{N_E}}}} \right|]$, ${h_2}\left( {{P_{p,i}},{P_{s,i}},{P_{z,i}}} \right) = {\log _2}$ $({{b_i}{P_{p,i}}\! +\! \sigma _{c,i}^2} ) + \rm{E}[{\log _2}\left| {{{\bf{c}}_i}{P_{p,i}}} \right.\! +\! {{\bf{f}}_i}{P_{s,i}}\! +\! {{\bf{g}}_i}{P_{z,i}}$ $\! + \!\left. {\sigma _{e,i}^2{{\bf{I}}_{{N_E}}}} \right|]$, ${r_1}\left( {{P_{p,i}},{P_{s,i}},{P_{z,i}}} \right) = {\log _2}\left( {{a_i}{P_{p,i}} + {d_i}{P_{s,i}} + \sigma _{p,i}^2} \right) + \rm{E}[{\log _2}|$ ${{{\bf{f}}_i}{P_{s,i}} + {{\bf{g}}_i}{P_{z,i}} + \sigma _{e,i}^2{{\bf{I}}_{{N_E}}}}|]$ and ${r_2}\left( {{P_{p,i}},{P_{s,i}},{P_{z,i}}} \right) = {\log _2}($ ${{d_i}{P_{s,i}} + \sigma _{p,i}^2}) + \rm{E}[{\log _2}\left| {{{\bf{c}}_i}{P_{p,i}} + {{\bf{f}}_i}{P_{s,i}} + {{\bf{g}}_i}{P_{z,i}} + \sigma _{e,i}^2{{\bf{I}}_{{N_E}}}} \right|]$. The gradients of ${h_2}({P_{p,i}},{P_{s,i}},{P_{z,i}})$ and ${r_2}( {{P_{p,i}},{P_{s,i}},{P_{z,i}}})$ are respectively written as
\begin{equation}\tag{27}
\begin{split}
&d{h_2}\left( {{P_{p,i}},{P_{s,i}},{P_{z,i}}} \right) = \frac{{{b_i}}}{{\left( {{b_i}{P_{p,i}} + \sigma _{c,i}^2} \right)\ln 2}}d{P_{p,i}}\\
& + \frac{{{\rm{E}}\left\{{\rm{Tr}}\left[ {{{\bf{c}}_i}{{\left( {{{\bf{c}}_i}{P_{p,i}} + {{\bf{f}}_i}{P_{s,i}} + {{\bf{g}}_i}{P_{z,i}} + \sigma _{e,i}^2{{\bf{I}}_{{N_E}}}} \right)}^{ - 1}}d{P_{p,i}}} \right]\right\} }}{{\ln 2}}\\
& + \frac{{{\rm{E}}\left\{{\rm{Tr}}\left[ {{{\bf{f}}_i}{{\left( {{{\bf{c}}_i}{P_{p,i}} + {{\bf{f}}_i}{P_{s,i}} + {{\bf{g}}_i}{P_{z,i}} + \sigma _{e,i}^2{{\bf{I}}_{{N_E}}}} \right)}^{ - 1}}d{P_{s,i}}} \right]\right\}}}{{\ln 2}}\\
& + \frac{{{\rm{E}}\left\{{\rm{Tr}}\left[ {{{\bf{g}}_i}{{\left( {{{\bf{c}}_i}{P_{p,i}} + {{\bf{f}}_i}{P_{s,i}} + {{\bf{g}}_i}{P_{z,i}} + \sigma _{e,i}^2{{\bf{I}}_{{N_E}}}} \right)}^{ - 1}}d{P_{z,i}}} \right]\right\}}}{{\ln 2}},
\end{split}
\end{equation}
and
\begin{equation}\tag{28}
\begin{split}
&d{r_2}\left( {{P_{p,i}},{P_{s,i}},{P_{z,i}}} \right) = \frac{{{d_i}}}{{\left( {{d_i}{P_{s,i}} + \sigma _{p,i}^2} \right)\ln 2}}d{P_{s,i}} \\
& + \frac{{{\rm{E}}\left\{{\rm{Tr}}\left[ {{{\bf{c}}_i}{{\left( {{{\bf{c}}_i}{P_{p,i}} + {{\bf{f}}_i}{P_{s,i}} + {{\bf{g}}_i}{P_{z,i}} + \sigma _{e,i}^2{{\bf{I}}_{{N_E}}}} \right)}^{ - 1}}d{P_{p,i}}} \right]\right\}}}{{\ln 2}}\\
& + \frac{{{\rm{E}}\left\{{\rm{Tr}}\left[ {{{\bf{f}}_i}{{\left( {{{\bf{c}}_i}{P_{p,i}} + {{\bf{f}}_i}{P_{s,i}} + {{\bf{g}}_i}{P_{z,i}} + \sigma _{e,i}^2{{\bf{I}}_{{N_E}}}} \right)}^{ - 1}}d{P_{s,i}}} \right]\right\}}}{{\ln 2}}\\
& + \frac{{{\rm{E}}\left\{{\rm{Tr}}\left[ {{{\bf{g}}_i}{{\left( {{{\bf{c}}_i}{P_{p,i}} + {{\bf{f}}_i}{P_{s,i}} + {{\bf{g}}_i}{P_{z,i}} + \sigma _{e,i}^2{{\bf{I}}_{{N_E}}}} \right)}^{ - 1}}d{P_{z,i}}} \right]\right\}}}{{\ln 2}}.
\end{split}
\end{equation}
Then, assuming $({\widetilde P_{p,i}},{\widetilde P_{s,i}},{\widetilde P_{z,i}})$ is a feasible solution of ${h_2}\left( {{P_{p,i}},{P_{s,i}},{P_{z,i}}} \right)$ and ${r_2}\left( {{P_{p,i}},{P_{s,i}},{P_{z,i}}} \right)$, the first-order Taylor series expansions of ${h_2}\left( {{P_{p,i}},{P_{s,i}},{P_{z,i}}} \right)$ and ${r_2}\left( {{P_{p,i}},{P_{s,i}},{P_{z,i}}} \right)$ can be obtained as
\begin{equation}\tag{29}
\begin{split}
&{h_2}\left( {{P_{p,i}},{P_{s,i}},{P_{z,i}}} \right) \le {h_2}( {{\widetilde P_{p,i}},{\widetilde P_{s,i}},{\widetilde P_{z,i}}})\! +\! \frac{{{b_i}({P_{p,i}} - {{\widetilde P_{p,i}}})}}{{({b_i}{{\widetilde P_{p,i}}} + \sigma _{c,i}^2)\ln 2}}  \\
 &\!+\! \frac{{{\rm E}\!\left\{ \!{{\rm{Tr}}\!\left[ {{{\bf{c}}_i}{{( {{{\bf{c}}_i}{{\widetilde P_{p,i}}} \!+ \!{{\bf{f}}_i}{{\widetilde P_{s,i}}} \!+ \!{{\bf{g}}_i}{{\widetilde P_{z,i}}} \!+\! \sigma _{e,i}^2{{\bf{I}}_{{N_E}}}} )}^{ - 1}}\!( {{P_{p,i}}\! - \!{{\widetilde P_{p,i}}}} )} \right]} \!\right\}}}{{\ln 2}}\\
 &\!+\! \frac{{{\rm E}\!\left\{ \!{{\rm{Tr}}\!\left[ {{{\bf{f}}_i}{{( {{{\bf{c}}_i}{{\widetilde P_{p,i}}} \!+ \!{{\bf{f}}_i}{{\widetilde P_{s,i}}} \!+ \!{{\bf{g}}_i}{{\widetilde P_{z,i}}} \!+\! \sigma _{e,i}^2{{\bf{I}}_{{N_E}}}} )}^{ - 1}}\!( {{P_{s,i}}\! - \!{{\widetilde P_{s,i}}}} )} \right]} \!\right\}}}{{\ln 2}}\\
 &\!+\! \frac{{{\rm E}\!\left\{ \!{{\rm{Tr}}\!\left[ {{{\bf{g}}_i}{{( {{{\bf{c}}_i}{{\widetilde P_{p,i}}} \!+ \!{{\bf{f}}_i}{{\widetilde P_{s,i}}} \!+ \!{{\bf{g}}_i}{{\widetilde P_{z,i}}} \!+\! \sigma _{e,i}^2{{\bf{I}}_{{N_E}}}} )}^{ - 1}}\!( {{P_{z,i}}\! - \!{{\widetilde P_{z,i}}}} )} \right]} \!\right\}}}{{\ln 2}},
\end{split}
\end{equation}
and
\begin{equation}\tag{30}
\begin{split}
&{r_2}\left( {{P_{p,i}},{P_{s,i}},{P_{z,i}}} \right) \le {r_2}( {{\widetilde P_{p,i}},{\widetilde P_{s,i}},{\widetilde P_{z,i}}})\! +\! \frac{{{d_i}({P_{s,i}} - {{\widetilde P_{s,i}}})}}{{({b_i}{{\widetilde P_{s,i}}} + \sigma _{p,i}^2)\ln 2}}  \\
 &\!+\! \frac{{{\rm E}\!\left\{ \!{{\rm{Tr}}\!\left[ {{{\bf{c}}_i}{{( {{{\bf{c}}_i}{{\widetilde P_{p,i}}} \!+ \!{{\bf{f}}_i}{{\widetilde P_{s,i}}} \!+ \!{{\bf{g}}_i}{{\widetilde P_{z,i}}} \!+\! \sigma _{e,i}^2{{\bf{I}}_{{N_E}}}} )}^{ - 1}}\!( {{P_{p,i}}\! - \!{{\widetilde P_{p,i}}}} )} \right]} \!\right\}}}{{\ln 2}}\\
 &\!+\! \frac{{{\rm E}\!\left\{ \!{{\rm{Tr}}\!\left[ {{{\bf{f}}_i}{{( {{{\bf{c}}_i}{{\widetilde P_{p,i}}} \!+ \!{{\bf{f}}_i}{{\widetilde P_{s,i}}} \!+ \!{{\bf{g}}_i}{{\widetilde P_{z,i}}} \!+\! \sigma _{e,i}^2{{\bf{I}}_{{N_E}}}} )}^{ - 1}}\!( {{P_{s,i}}\! - \!{{\widetilde P_{s,i}}}} )} \right]} \!\right\}}}{{\ln 2}}\\
 &\!+\! \frac{{{\rm E}\!\left\{ \!{{\rm{Tr}}\!\left[ {{{\bf{g}}_i}{{( {{{\bf{c}}_i}{{\widetilde P_{p,i}}} \!+ \!{{\bf{f}}_i}{{\widetilde P_{s,i}}} \!+ \!{{\bf{g}}_i}{{\widetilde P_{z,i}}} \!+\! \sigma _{e,i}^2{{\bf{I}}_{{N_E}}}} )}^{ - 1}}\!( {{P_{z,i}}\! - \!{{\widetilde P_{z,i}}}} )} \right]} \!\right\}}}{{\ln 2}}.
\end{split}
\end{equation}

\begin{table*}[t]
\begin{equation}\renewcommand\theequation{25}
\begin{split}
\mathop {\max }\limits_{{{\bf{P}}_p},{{\bf{P}}_s},{{\bf{P}}_z}} {\kern 1pt} {\kern 1pt} {\kern 1pt} {\kern 1pt} {\kern 1pt} &{\varphi _{{\rm{SEE}}}} = \frac{{\sum\limits_{i = 1}^I {\left\{ {{{\log }_2}\left( {1 + \frac{{{e_i}{P_{s,i}}}}{{{b_i}{P_{p,i}} + \sigma _{c,i}^2}}} \right) - {\rm{E}}\left[ {{{\log }_2}\frac{{\left| {{{\bf{c}}_i}{P_{p,i}} + {{\bf{f}}_i}{P_{s,i}} + {{\bf{g}}_i}{P_{z,i}} + \sigma _{e,i}^2{{\bf{I}}_{{N_E}}}} \right|}}{{\left| {{{\bf{c}}_i}{P_{p,i}} + {{\bf{g}}_i}{P_{z,i}} + \sigma _{e,i}^2{{\bf{I}}_{{N_E}}}} \right|}}} \right]} \right\}} }}{{\sum\limits_{i = 1}^I {\left( {{P_{s,i}} + {P_{z,i}}} \right) + {P_b}} }}\\
{\kern 1pt} {\kern 1pt} {\kern 1pt} {\kern 1pt} {\kern 1pt} s.t.{\kern 1pt} {\kern 1pt} {\kern 1pt} {\kern 1pt} {\kern 1pt} {\kern 1pt} {\kern 1pt} {\kern 1pt} {\kern 1pt} {\kern 1pt} {\kern 1pt} {\kern 1pt}& C1:{\log _2}\left( {1 + \frac{{{e_i}{P_{s,i}}}}{{{b_i}{P_{p,i}} + \sigma _{c,i}^2}}} \right) - {\rm{E}}\left[ {{{\log }_2}\frac{{\left| {{{\bf{c}}_i}{P_{p,i}} + {{\bf{f}}_i}{P_{s,i}} + {{\bf{g}}_i}{P_{z,i}} + \sigma _{e,i}^2{{\bf{I}}_{{N_E}}}} \right|}}{{\left| {{{\bf{c}}_i}{P_{p,i}} + {{\bf{g}}_i}{P_{z,i}} + \sigma _{e,i}^2{{\bf{I}}_{{N_E}}}} \right|}}} \right] \ge R_{CU}^{\min },{\kern 1pt} {\kern 1pt} {\kern 1pt} {\kern 1pt} {\kern 1pt} {\kern 1pt} {\kern 1pt} \forall i,\\
{\kern 1pt} {\kern 1pt} {\kern 1pt} {\kern 1pt} {\kern 1pt} {\kern 1pt} {\kern 1pt} {\kern 1pt} {\kern 1pt} {\kern 1pt} {\kern 1pt} {\kern 1pt} {\kern 1pt} {\kern 1pt} {\kern 1pt} {\kern 1pt} {\kern 1pt} {\kern 1pt} {\kern 1pt} {\kern 1pt} {\kern 1pt} {\kern 1pt} {\kern 1pt} {\kern 1pt} {\kern 1pt} {\kern 1pt} {\kern 1pt} {\kern 1pt} {\kern 1pt}& C2:{\log _2}\left( {1 + \frac{{{a_i}{P_{p,i}}}}{{{d_i}{P_{s,i}} + \sigma _{p,i}^2}}} \right) - {\rm{E}}\left[ {{{\log }_2}\frac{{\left| {{{\bf{c}}_i}{P_{p,i}} + {{\bf{f}}_i}{P_{s,i}} + {{\bf{g}}_i}{P_{z,i}} + \sigma _{e,i}^2{{\bf{I}}_{{N_E}}}} \right|}}{{\left| {{{\bf{f}}_i}{P_{s,i}} + {{\bf{g}}_i}{P_{z,i}} + \sigma _{e,i}^2{{\bf{I}}_{{N_E}}}} \right|}}} \right] \ge R_{PU}^{\min },{\kern 1pt} {\kern 1pt} {\kern 1pt} {\kern 1pt} \forall i,\\
{\kern 1pt} {\kern 1pt} {\kern 1pt} {\kern 1pt} {\kern 1pt} {\kern 1pt} {\kern 1pt} {\kern 1pt} {\kern 1pt} {\kern 1pt} {\kern 1pt} {\kern 1pt} {\kern 1pt} {\kern 1pt} {\kern 1pt} {\kern 1pt} {\kern 1pt} {\kern 1pt} {\kern 1pt} {\kern 1pt} {\kern 1pt} {\kern 1pt} {\kern 1pt} {\kern 1pt} {\kern 1pt} {\kern 1pt} {\kern 1pt} {\kern 1pt} {\kern 1pt}& C3:\sum\limits_{i = 1}^I {{P_{p,i}}}  \le P_{{\rm{PBS}}}^{{\rm{total}}},\\
{\kern 1pt} {\kern 1pt} {\kern 1pt} {\kern 1pt} {\kern 1pt} {\kern 1pt} {\kern 1pt} {\kern 1pt} {\kern 1pt} {\kern 1pt} {\kern 1pt} {\kern 1pt} {\kern 1pt} {\kern 1pt} {\kern 1pt} {\kern 1pt} {\kern 1pt} {\kern 1pt} {\kern 1pt} {\kern 1pt} {\kern 1pt} {\kern 1pt} {\kern 1pt} {\kern 1pt} {\kern 1pt} {\kern 1pt} {\kern 1pt} {\kern 1pt} {\kern 1pt} {\kern 1pt}& C4:\sum\limits_{i = 1}^I {\left( {{P_{s,i}} + {P_{z,i}}} \right) \le P_{{\rm{CBS}}}^{{\rm{total}}},}
\end{split}
\end{equation}
\hrule
\end{table*}

\begin{table*}[t]
\begin{equation}\renewcommand\theequation{31}
\begin{split}
&\mathop {\max }\limits_{{{\bf{P}}_p},{{\bf{P}}_s},{{\bf{P}}_z}} \sum\limits_{i = 1}^I {\left\{ {h_1}\left( {{P_{p,i}},{P_{s,i}},{P_{z,i}}} \right) \right.}\! -\!{h_2}( {{{\widetilde P}_{p,i}},{{\widetilde P}_{s,i}},{{\widetilde P}_{z,i}}})\! -\! \frac{{{b_i}({P_{p,i}} - {{\widetilde P}_{p,i}})}}{{({b_i}{{\widetilde P}_{p,i}} + \sigma _{c,i}^2)\ln 2}}\! -\! \frac{{{\rm E}\left\{ {{\rm{Tr}}\left[ {{{\bf{c}}_i}{{( {\bf{\widetilde \Omega }}_i)}^{ - 1}}( {{P_{p,i}}\! -\! {{\widetilde P}_{p,i}}})} \right]} \right\}}}{{\ln 2}}\! -\! \frac{{{\rm E}\left\{ {{\rm{Tr}}\left[ {{{\bf{f}}_i}{{({\bf{\widetilde \Omega }}_i)}^{ - 1}}( {{P_{s,i}}\! -\! {{\widetilde P}_{s,i}}} )} \right]} \right\}}}{{\ln 2}} \\
 &\qquad\qquad - \!\!\left. {\frac{{{\rm E}\left\{ {{\rm{Tr}}\left[ {{{\bf{g}}_i}{{( {\bf{\widetilde \Omega }}_i)}^{ - 1}}( {{P_{z,i}} - {{\widetilde P}_{z,i}}} )} \right]} \right\}}}{{\ln 2}}} \!\right\} -{\varphi _{{\rm{SEE}}}}\left[ {\sum\limits_{i = 1}^I {\left( {{P_{s,i}} + {P_{z,i}}} \right) + {P_b}} } \right]\\
&s.t.{\kern 1pt} C1:{h_1}\left( {{P_{p,i}},{P_{s,i}},{P_{z,i}}} \right) - {h_2}( {{{\widetilde P}_{p,i}},{{\widetilde P}_{s,i}},{{\widetilde P}_{z,i}}} ) - \frac{{{b_i}({P_{p,i}} - {{\widetilde P}_{p,i}})}}{{({b_i}{{\widetilde P}_{p,i}} + \sigma _{c,i}^2)\ln 2}} - \frac{{{\rm E}\left\{ {{\rm{Tr}}\left[ {{{\bf{c}}_i}{{({\bf{\widetilde \Omega }}_i)}^{ - 1}}( {{P_{p,i}} - {{\widetilde P}_{p,i}}} )} \right]} \right\}}}{{\ln 2}}\! -\! \frac{{{\rm E}\left\{ {{\rm{Tr}}\left[ {{{\bf{f}}_i}{{( {\bf{\widetilde \Omega }}_i )}^{ - 1}}( {{P_{s,i}} - {{\widetilde P}_{s,i}}} )} \right]} \right\}}}{{\ln 2}}\\
 & \qquad\qquad - \frac{{{\rm E}\left\{ {{\rm{Tr}}\left[ {{{\bf{g}}_i}{{({\bf{\widetilde \Omega }}_i )}^{ - 1}}( {{P_{z,i}} - {{\widetilde P}_{z,i}}} )} \right]} \right\}}}{{\ln 2}} \ge R_{CU}^{\min },\quad \forall i,\\
 &\qquad\!\! C2:{r_1}\left( {{P_{p,i}},{P_{s,i}},{P_{z,i}}} \right) - {r_2}( {{{\widetilde P}_{p,i}},{{\widetilde P}_{s,i}},{{\widetilde P}_{z,i}}} ) - \frac{{{d_i}({P_{s,i}} - {{\widetilde P}_{s,i}})}}{{({d_i}{{\widetilde P}_{s,i}} + \sigma _{p,i}^2)\ln 2}} - \frac{{{\rm E}\left\{ {{\rm{Tr}}\left[ {{{\bf{c}}_i}{{({\bf{\widetilde \Omega }}_i)}^{ - 1}}( {{P_{p,i}} - {{\widetilde P}_{p,i}}} )} \right]} \right\}}}{{\ln 2}} - \frac{{{\rm E}\left\{ {{\rm{Tr}}\left[ {{{\bf{f}}_i}{{( {\bf{\widetilde \Omega }}_i )}^{ - 1}}( {{P_{s,i}} - {{\widetilde P}_{s,i}}} )} \right]} \right\}}}{{\ln 2}}\\
&\qquad\qquad - \frac{{{\rm E}\left\{ {{\rm{Tr}}\left[ {{{\bf{g}}_i}{{( {\bf{\widetilde \Omega }}_i )}^{ - 1}}( {{P_{z,i}} - {{\widetilde P}_{z,i}}} )} \right]} \right\}}}{{\ln 2}} \ge R_{PU}^{\min },\quad \forall i,\\
 &\qquad\!\! C3,\quad\!\!\! C4.
\end{split}
\end{equation}
\hrule
\end{table*}

\begin{table*}[t]
\begin{equation}\renewcommand\theequation{32}
\begin{split}
&(\widetilde {\bf{P}}_p^{n + 1},\widetilde {\bf{P}}_s^{n + 1},\widetilde {\bf{P}}_z^{n + 1})=\arg\mathop {\max }\limits_{{{\bf{P}}_p},{{\bf{P}}_s},{{\bf{P}}_z}} \sum\limits_{i = 1}^I {\left\{ {h_1}\left( {{P_{p,i}},{P_{s,i}},{P_{z,i}}} \right) \right.}\! -\!{h_2}( {{{\widetilde P}_{p,i}^n},{{\widetilde P}_{s,i}^n},{{\widetilde P}_{z,i}^n}})\! -\! \frac{{{b_i}({P_{p,i}} - {{\widetilde P}_{p,i}^n})}}{{({b_i}{{\widetilde P}_{p,i}^n} + \sigma _{c,i}^2)\ln 2}}\! -\! \frac{{{\rm E}\left\{ {{\rm{Tr}}\left[ {{{\bf{c}}_i}{{( {\bf{\widetilde \Omega }^n}_i)}^{ - 1}}( {{P_{p,i}}\! -\! {{\widetilde P}_{p,i}^n}})} \right]} \right\}}}{{\ln 2}}\\
&\qquad\qquad -\! \frac{{{\rm E}\left\{ {{\rm{Tr}}\left[ {{{\bf{f}}_i}{{({\bf{\widetilde \Omega }^n}_i)}^{ - 1}}( {{P_{s,i}}\! -\! {{\widetilde P}_{s,i}^n}} )} \right]} \right\}}}{{\ln 2}} - \!\!\left. {\frac{{{\rm E}\left\{ {{\rm{Tr}}\left[ {{{\bf{g}}_i}{{( {\bf{\widetilde \Omega }^n}_i)}^{ - 1}}( {{P_{z,i}} - {{\widetilde P}_{z,i}^n}} )} \right]} \right\}}}{{\ln 2}}} \!\right\} - {\varphi _{{\rm{SEE}}}}\left[ {\sum\limits_{i = 1}^I {\left( {{P_{s,i}} + {P_{z,i}}} \right) + {P_b}} } \right]\\
&s.t.{\kern 2pt} C1:{h_1}\left( {{P_{p,i}},{P_{s,i}},{P_{z,i}}} \right) - {h_2}( {{\widetilde P_{p,i}^n},{\widetilde P_{s,i}^n},{\widetilde P_{z,i}^n}} ) -\frac{{{b_i}({P_{p,i}} \!-\! \widetilde P_{p,i}^n)}}{{({b_i}\widetilde P_{p,i}^n \!+\! \sigma _{c,i}^2)\!\ln2}} - \frac{{{\rm E}\!\left\{ \!{{\rm{Tr}}\!\left[ {{{\bf{c}}_i}{{({\bf{\tilde \Omega }}_i^n)}^{ - 1}}\!({P_{p,i}}\! -\! \widetilde P_{p,i}^n)} \right]} \!\right\}}}{{\ln 2}} \!-\! \frac{{{\rm E}\!\left\{ \!{{\rm{Tr}}\!\left[ {{{\bf{f}}_i}{{({\bf{\tilde \Omega }}_i^n)}^{ - 1}}\!({P_{s,i}}\! -\! \widetilde P_{s,i}^n)} \right]} \!\right\}}}{{\ln 2}} \\
&\qquad\qquad-\! \frac{{{\rm E}\!\left\{ \!{{\rm{Tr}}\!\left[ {{{\bf{g}}_i}{{({\bf{\tilde \Omega }}_i^n)}^{ - 1}}\!({P_{z,i}}\! -\! \widetilde P_{z,i}^n)} \right]} \!\right\}}}{{\ln 2}}\ge R_{CU}^{\min },{\kern 1pt} {\kern 1pt} {\kern 1pt} {\kern 1pt} {\kern 1pt} {\kern 1pt} {\kern 1pt} \forall i,\\
& \qquad\!\! C2:{r_1}\left( {{P_{p,i}},{P_{s,i}},{P_{z,i}}} \right) - {r_2}( {{\widetilde P_{p,i}^n},{\widetilde P_{s,i}^n},{\widetilde P_{z,i}^n}} )-\frac{{{d_i}({P_{s,i}} \!-\! \widetilde P_{s,i}^n)}}{{({d_i}\widetilde P_{s,i}^n \!+\! \sigma _{p,i}^2)\!\ln2}}\!-\! \frac{{{\rm E}\!\left\{ \!{{\rm{Tr}}\!\left[ {{{\bf{c}}_i}{{({\bf{\tilde \Omega }}_i^n)}^{ - 1}}\!({P_{p,i}}\! -\! \widetilde P_{p,i}^n)} \right]} \!\right\}}}{{\ln 2}}\!-\! \frac{{{\rm E}\!\left\{ \!{{\rm{Tr}}\!\left[ {{{\bf{f}}_i}{{({\bf{\tilde \Omega }}_i^n)}^{ - 1}}\!({P_{s,i}}\! -\! \widetilde P_{s,i}^n)} \right]} \!\right\}}}{{\ln 2}}\\
&\qquad\qquad\!-\! \frac{{{\rm E}\!\left\{ \!{{\rm{Tr}}\!\left[ {{{\bf{g}}_i}{{({\bf{\tilde \Omega }}_i^n)}^{ - 1}}\!({P_{z,i}}\! -\! \widetilde P_{z,i}^n)} \right]} \!\right\}}}{{\ln 2}}\ge R_{PU}^{\min },{\kern 1pt} {\kern 1pt} {\kern 1pt} {\kern 1pt} {\kern 1pt} {\kern 1pt} {\kern 1pt} \forall i,\\
& \qquad\!\!C3,\quad\!\!\! C4.
\end{split}
\end{equation}
\hrule
\end{table*}

Denoting ${{{\bf{\widetilde \Omega }}_i}} = {{\bf{c}}_i}{{\widetilde P}_{p,i}} + {{\bf{f}}_i}{{\widetilde P}_{s,i}} + {{\bf{g}}_i}{{\widetilde P}_{z,i}} + \sigma _{e,i}^2{{\bf{I}}_{{N_E}}}$ and according to Section III-C, we employ the D.C. approximation method [37] to transform (26) into an approximate convex problem (31) at top of the next page. As a result, assuming that $(\widetilde P_{p,i}^n,\widetilde P_{s,i}^n,\widetilde P_{z,i}^n)$ and $(\widetilde P_{p,i}^{n + 1},\widetilde P_{s,i}^{n + 1},\widetilde P_{z,i}^{n + 1})$ are the optimal solutions to (31) at iterations $n$ and $n+1$, and letting ${{{\bf{\widetilde \Omega }}_i^{n}}} = {{\bf{c}}_i}{{\widetilde P}_{p,i}^{n}} + {{\bf{f}}_i}{{\widetilde P}_{s,i}^{n}} + {{\bf{g}}_i}{{\widetilde P}_{z,i}^{n}} + \sigma _{e,i}^2{{\bf{I}}_{{N_E}}}$, the solution of (26) can be obtained through the iterative procedure at (32).
According to Appendix B, the convergence of the iterative procedure can be guaranteed. Then, the optimization problem (31) can be easily solved by CVX [40]. Finally, a two-tier iterative $\varepsilon$-optimal power allocation algorithm for SCSI-SEEM scheme is summarized in Table II. The ${\varphi _{\rm{SEE}}}$ satisfying $h\left( {{\varphi _{\rm{SEE}}}} \right) = 0$ is found with the help of Dinkelbach's method [41] at the outer tier, meanwhile, the solution is achieved for a given ${\varphi _{\rm{SEE}}}$ at the inner tier.

In addition, the computational complexity of proposed SCSI-SEEM scheme is determined by the number of iterations, variable size and the number of constraints at the outer and inner tiers. The iterations excluding convex programming can be given by $O\left( {\log \left( {\varphi _{\textrm{SEE}}^{up}/\varepsilon } \right)\log \left( {\phi _{\textrm{SEE}}^{up}/\varepsilon } \right)} \right)$, where $\varphi _{\textrm{SEE}}^{up} = ( {\frac{{\max ({e_i})P_{\textrm{CBS}}^{\textrm{total}}}}{{\Delta f{N_0}\ln 2}}} )/{P_b}$, $\phi_{\textrm{SEE}}^{up} = \frac{{\max ({e_i})P_{\textrm{CBS}}^{\textrm{total}}}}{{\Delta f{N_0}\ln 2}}$, and $\varepsilon$ is the tolerance level. Since the problem (32) has $3I$ variables, we need at most $O((3I)^{3.5}\log (1/\varepsilon ))$ calculations at each inner iteration [42]. Thus, the overall computational complexity of the SCSI-SEEM scheme can be given by
\begin{equation}\tag{33}
O\left( {\log \left( {\frac{1}{\varepsilon }} \right)\log \left( {\frac{{\varphi _{{\rm{SEE}}}^{up}}}{\varepsilon }} \right)\log \left( {\frac{{\phi_{\rm{SEE}}^{up}}}{\varepsilon }} \right){{\left( {3I} \right)}^{3.5}}} \right).
\end{equation}

\vspace{0.05in}
\begin{table}[!h]
\caption{Two-tier Iterative $\varepsilon$-optimal Power Allocation Algorithm for SCSI-SEEM Scheme}
\begin{center}
\begin{tabular}{l}
\hline
Algorithm 2: Two-tier Iterative $\varepsilon$-optimal Power Allocation Algorithm.\\
\hline
\textbf{Function}
\emph{Outer{\_} Iteration}\\
\small
Step 1: Initialize the maximum number of iterations ${m_{\textrm{max} }}$, ${n_{\textrm{max} }}$\\
~~~~~~~~and the maximum tolerance $\varepsilon$.\\
\small
Step 2: Set maximum SEE $\varphi _{\textrm{SEE}}^0 = 0$ and iteration index
 $m=0$.\\
\small
Step 3: Call
\textbf{Function}
\emph{Inner{\_} Iteration} with $\varphi _{\textrm{SEE}}^m$ to obtain the\\
{\kern 25pt}$\varepsilon$-optimal solution $( {\bf{P}}_{p}^n,{\bf{P}}_{s}^n,{\bf{P}}_{z}^n)$.\\
\small
Step 4: Update $\varphi _{{\rm{SEE}}}^{m + 1}$ = \\
$\quad \frac{{\sum\limits_{i = 1}^I\!\! {\left\{{{{\log }_2}\left( {1 + \frac{{{e_i}P_{s,i}^n}}{{{b_i}P_{p,i}^n + \sigma _{c,i}^2}}} \right)\! - {\rm{E}}\left[ {{\log}_2}{\frac{{\left| {{{\bf{c}}_i}P_{p,i}^n + {{\bf{f}}_i}P_{s,i}^n + {{\bf{g}}_i}P_{z,i}^n + \sigma _{e,i}^2{{\bf{I}}_{{N_E}}}} \right|}}{{\left| {{{\bf{c}}_i}P_{p,i}^n + {{\bf{g}}_i}P_{z,i}^n + \sigma _{e,i}^2{{\bf{I}}_{{N_E}}}} \right|}}} \right]} \right\}} }}{{\sum\limits_{i = 1}^I {\left( {P_{s,i}^n + P_{z,i}^n} \right)}  + {P_b}}}$.\\
\small
Step 5: Set $m=m+1$.\\
\small
Step 6: \textbf{if} $\left| {\varphi _{\textrm{SEE}}^m - \varphi _{\textrm{SEE}}^{m - 1}} \right| \ge \varepsilon $ or $m \le {m_{\textrm{max} }}$ \\
\small
Step 7: \textbf{goto} Step 3.\\
\small
Step 8: \textbf{end if}\\
\small
Step 9: \textbf{return} ${\bf{P}}_p^{n}$, ${\bf{P}}_s^{n}$, ${\bf{P}}_z^{n}$.\\
\small
Step 10: Obtain the $\varepsilon$-optimal solution ${\bf{P}}_{p}^ *  = {\bf{P}}_{p}^n$, ${\bf{P}}_{s}^ *  = {\bf{P}}_{s}^n$ \\
{\kern 30pt} and ${\bf{P}}_{z}^ *  = {\bf{P}}_{z}^n$ for problem (25).\\
\small
\textbf{end} \\
\small
\textbf{Function}
\emph{Inner{\_} Iteration}  $(\varphi _{\textrm{SEE}})$\\
\small
Step 11: Initialize $({\bf{P}} _{p}^0, {\bf{P}} _{s}^0, {\bf{P}} _{z}^0) = (0,0,0)$ and ${h^0} = 0$.\\
\small
Step 12: Set $n = 0$.\\
\small
Step 13: Find the $\varepsilon$-optimal solution $( {{\bf{P}}_{p}^{n+1}},{{\bf{P}}_{s}^{n+1}}, {{\bf{P}}_{z}^{n+1}} )$ of (30)\\
{\kern 30pt} for given $({{\bf{P}}_{p}^{n}},{{\bf{P}}_{s}^{n}},{{\bf{P}}_{z}^{n}})$ and ${\varphi _{\textrm{SEE}}^m}$ by using CVX.\\
\small
Step 14: Compute\\
${h^{n + 1}} = \sum\limits_{i = 1}^I {\left[ {{h_1}\left( {P_{p,i}^{n + 1},P_{s,i}^{n + 1},P_{z,i}^{n + 1}} \right) - {h_2}\left( {P_{p,i}^{n + 1},P_{s,i}^{n + 1},P_{z,i}^{n + 1}} \right)} \right]}  $\\
\qquad\qquad $- {\varphi _{{\rm{SEE}}}^{m}}\left[ {\sum\limits_{i = 1}^I {( {{P_{s,i}^{n+1}} + {P_{z,i}^{n+1}}} ) + {P_b}} } \right]$ \\
\small
Step 15: Set $n=n+1$.\\
\small
Step 16: \textbf{if} $\left| {{h^n} - {h^{n - 1}}} \right| \ge \varepsilon $ or $n \le {n_{\textrm{max} }}$  \\
\small
Step 17: \textbf{goto} Step 13.\\
\small
Step 18: \textbf{end if}\\
\small
Step 19: \textbf{return} ${{\bf{P}}_{p}^{n}}$, ${{\bf{P}}_{s}^{n}}$, ${{\bf{P}}_{z}^{n}}$.\\
\small
\textbf{end}\\
\hline
\end{tabular}
\end{center}
\end{table}

\section{Simulation Results}
In this section, we present numerical results to evaluate the performance of our proposed schemes. The simulation parameters can be found in Table III. All simulation results were averaged over 100 random channel realizations.

\begin{table}[!htp]
\caption{System Parameters}
\centering
\begin{tabular}{|c|c|}
\hline
Parameters &  Values\\
\hline
Path loss model, ${\log _{10}}\left( \vartheta  \right)$ & $ - 34.5 - 38{\log _{10}}(d[m])$ \\
\hline
SR threshold for PU, $R _{\rm{PU}}^{\rm{min}}$ & 0 bit/s/Hz \\
\hline
SR threshold for CU, $R _{\rm{CU}}^{\rm{min}}$ & 0 bit/s/Hz \\
\hline
Corresponding distance, $d$ & $500\quad\!\!\!\!\!\textrm{m}$ \\
\hline
Bandwidth, $\Delta f$ & 10 MHz \\
\hline
Noise spectral density, ${N_0}$ & -174 dBm/Hz \\
\hline
Basic power consumption of CBS, ${P_b}$ & 40 dBm \\
\hline
Maximum iteration, ${i_{\textrm{max} }}$ & 100 \\
\hline
Convergence threshold, $\varepsilon $ & ${10^{ - 3}}$ \\
\hline
\end{tabular}
\end{table}

{\begin{figure}[!h]
  \centering
 {\includegraphics[scale=0.6]{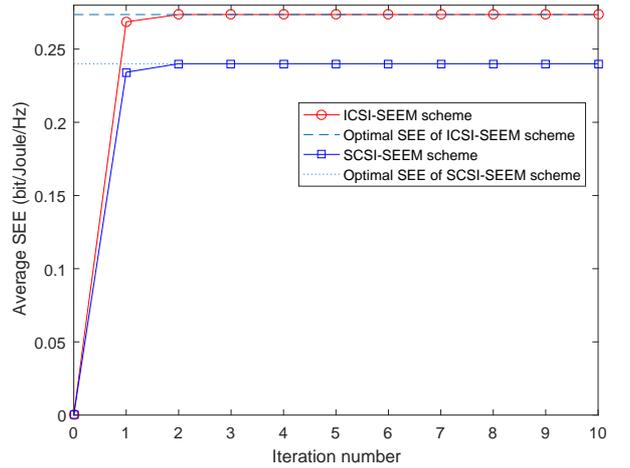}\\
  \caption{Covergence behavior of proposed algorithms for ICSI-SEEM and SCSI-SEEM schemes versus the number of iterations in terms of average SEE, with $I=8$, $N_P=N_C=4$, $N_E=3$, and the maximum transmit power of PBS and CBS, $P_{\textrm{PBS}}^{\textrm{total}}=30\textrm{dBm}$, $P_{\textrm{CBS}}^{\textrm{total}}=40\textrm{dBm}$.}\label{Fig2}}
\end{figure}

Fig. 2 presents the convergence behavior of proposed algorithms for ICSI-SEEM and SCSI-SEEM schemes versus the number of iterations in terms of average SEE, with $I=8$, $N_P=N_C=4$, $N_E=3$, and the maximum transmit power of PBS and CBS, $P_{\textrm{PBS}}^{\textrm{total}}=30\textrm{dBm}$, $P_{\textrm{CBS}}^{\textrm{total}}=40\textrm{dBm}$. As observed, the average SEE results obtained by proposed algorithms converge to the optimal SEE of ICSI-SEEM and SCSI-SEEM schemes respectively after sufficient iterations, which confirms that proposed algorithms are able to achieve the optimal solutions of ICSI-SEEM and SCSI-SEEM schemes by simply increasing the number of iterations.

\begin{figure}[!h]
  \centering
 {\includegraphics[scale=0.6]{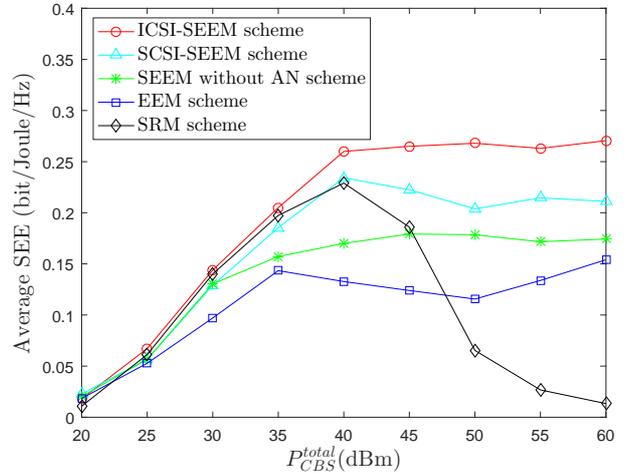}\\
  \caption{Average SEE versus maximum transmit power of CBS, $P_{\textrm{CBS}}^{\textrm{total}}$, with $I=8$, $N_P=N_C=4$, $N_E=3$, and the maximum transmit power of PBS, $P_{\textrm{PBS}}^{\textrm{total}}=30\textrm{dBm}$.}\label{Fig3}}
\end{figure}
Fig. 3 shows the average SEE results of proposed ICSI-SEEM and SCSI-SEEM
 as well as conventional SRM, EEM and SEEM without AN schemes versus the CBS transmit power constraint $P_{\textrm{CBS}}^{\textrm{total}}$ with $I=8$, $N_P=N_C=4$, $N_E=3$, and the maximum transmit power of PBS, $P_{\textrm{PBS}}^{\textrm{total}}=30\textrm{dBm}$. The average SEE performance of proposed ICSI-SEEM, SCSI-SEEM and conventional SRM schemes all improve with an increasing $P_{\textrm{CBS}}^{\textrm{total}}$ in the $20 - 40 \textrm{dBm}$ region of transmit power. This means that ICSI-SEEM, SCSI-SEEM and SRM schemes can obtain the maximum SEE with the full transmit power. Then, as $P_{\textrm{CBS}}^{\textrm{total}}$ continues to increase after $40 \textrm{dBm}$, the average SEE performance of proposed ICSI-SEEM and SCSI-SEEM schemes approach to a constant, while the SRM scheme begins to degrade in terms of its SEE performance. This is because that in the proposed ICSI-SEEM and SCSI-SEEM schemes, the power allocator would not consume more transmit power when the maximum SEE has been achieved. By contrast, in order to achieve a higher SR, the SRM scheme will continue to allocate more transmit power, which will result in the drop of the average SEE. In addition, as observed, the proposed ICSI-SEEM and SCSI-SEEM schemes significantly outperform the EEM scheme in terms of the average SEE, and ICSI-SEEM achieves a higher SEE than the SCSI-SEEM scheme. In the SEEM without AN scheme, CBS only transmits the confidential signal to the destination without considering AN, besides, the powers of CBS' and PBS' OFDM subcarriers are optimized with a given total power consumption for CBS and PBS, respectively. As can be seen from Fig. 2, the proposed ICSI-SEEM and SCSI-SEEM schemes achieve a higher SEE than SEEM without AN scheme, which indicates the advantage of AN to wiretap the ED.

\begin{figure}[!h]
 \centering
 {\includegraphics[scale=0.6]{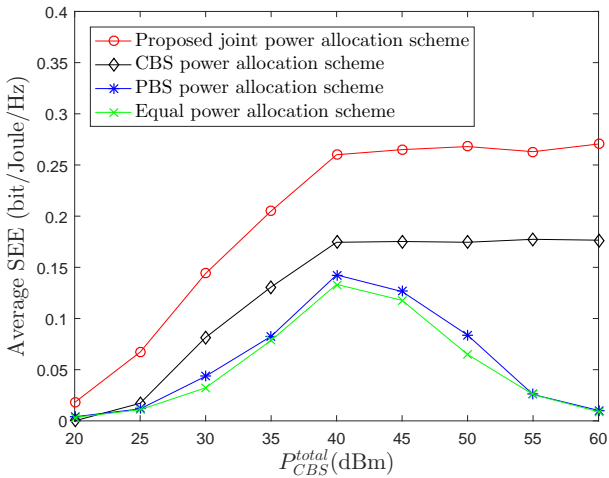}\\
  \caption{Average SEE versus maximum transmit power of CBS,  $P_{\textrm{CBS}}^{\textrm{total}}$, for different power allocation schemes with $I=8$, $N_P=N_C=4$, $N_E=3$, and the maximum transmit power of PBS, $P_{\textrm{PBS}}^{\textrm{total}}=30\textrm{dBm}$.}\label{Fig4}}
\end{figure}
Fig. 4 shows the average SEE versus maximum transmit power of CBS, $P_{\textrm{CBS}}^{\textrm{total}}$, for the proposed joint power allocation of PBS and CBS, pure power allocation of CBS' OFDM subcarriers (denoted by CBS power allocation for short), pure power allocation of PBS' OFDM subcarriers (called PBS power allocation) and equal power allocation schemes with $I=8$, $N_P=N_C=4$, $N_E=3$, and the maximum transmit power of PBS, $P_{\textrm{PBS}}^{\textrm{total}}=30\textrm{dBm}$. In the CBS power allocation scheme, the powers of CBS' OFDM subcarriers are optimized with a given total power consumption for CBS $P_{\textrm{CBS}}^{\textrm{total}}$ and a fixed power allocation is used for PBS' OFDM subcarriers, namely the power of each PBS' subcarrier is given by 10dBm. Similarly, the PBS power allocation scheme only considers the optimal power allocation for PBS' OFDM subcarriers with a constrained total power $P_{\textrm{PBS}}^{\textrm{total}}$, while the equal power allocation is used for CBS' subcarriers. Moreover, in the equal power allocation scheme, CBS' OFDM subcarriers are equally allocated with their respective total transmit power constraints while the PBS' OFDM subcarriers are allocated with fixed transmit power, namely, ${P_{s,i}} = {{P_{\textrm{CBS}}^{\textrm{total}}} \mathord{\left/
 {\vphantom {{P_{\textrm{CBS}}^{\textrm{total}}} {\left( {2I} \right)}}} \right.
 \kern-\nulldelimiterspace} {\left( {2I} \right)}}$, ${P_{z,i}} = {{P_{\textrm{CBS}}^{\textrm{total}}} \mathord{\left/
 {\vphantom {{P_{\textrm{CBS}}^{\textrm{total}}} {\left( {2I} \right)}}} \right.
 \kern-\nulldelimiterspace} {\left( {2I} \right)}}$ and ${P_{p,i}} = {\rm{10dBm}}$.

As can be seen from Fig. 4, the average SEE of proposed joint power allocation and CBS power allocation scheme approach a constant in the high CBS transmit power regime. This is because both the proposed joint power allocation and CBS power allocation schemes stop assuming more CBS transmit power when the maximal SEE is achieved. However, the PBS power allocation and equal power allocation schemes begin to drop in the regime of $P_{\textrm{CBS}}^{\textrm{total}} \ge 40\textrm{dBm}$. This is due to the fact that they allocate all the available CBS transmit power even without much secrecy rate improvement. On the other hand, the proposed joint power allocation scheme can achieve a higher average SEE than other power allocation methods, which indicates the superiority of proposed joint power allocation scheme.

\begin{figure}[!h]
 \centering
 {\includegraphics[scale=0.6]{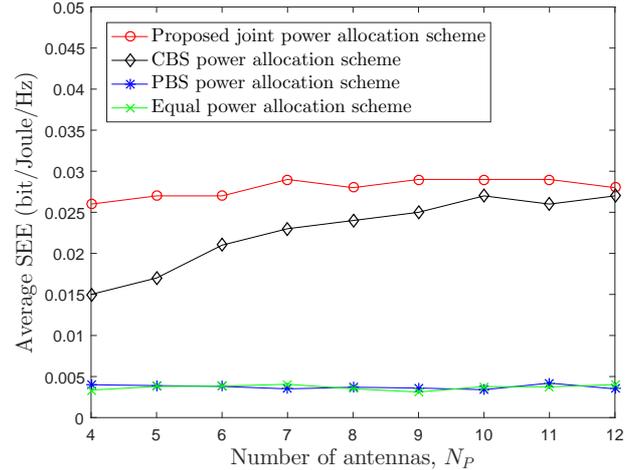}\\
  \caption{Average SEE versus the number of PBS' antennas, $N_P$, for different power allocation schemes with $I=8$, $N_C=4$, $N_E=3$, and the maximum transmit power of PBS and CBS, $P_{\textrm{PBS}}^{\textrm{total} }=30\textrm{dBm}$, $P_{\textrm{CBS}}^{\textrm{total} }=20\textrm{dBm}$.}\label{Fig5}}
\end{figure}
Fig. 5 illustrates the average SEE versus the number of PBS' antennas, $N_P$, for the proposed joint power allocation, CBS power allocation, PBS power allocation, and equal power allocation schemes with $I=8$, $N_C=4$, $N_E=3$, and the maximum transmit power of PBS and CBS, $P_{\textrm{PBS}}^{\textrm{total}}=30\textrm{dBm}$, $P_{\textrm{CBS}}^{\textrm{total}}=20\textrm{dBm}$. It can be observed that as $N_P$ increases, the CBS power allocation schemes begin to increase in terms of the average SEE, however, the proposed joint power allocation, PBS power allocation and equal power allocation methods converge to their respective SEE floors. This means that given sufficiently high number of PBS's antennas, the proposed joint power allocation, PBS power allocation and equal power allocation can sophisticatedly stop consuming additional power resources when the resultant secrecy rate improvement is marginal.

\begin{figure}[!h]
 \centering
 {\includegraphics[scale=0.6]{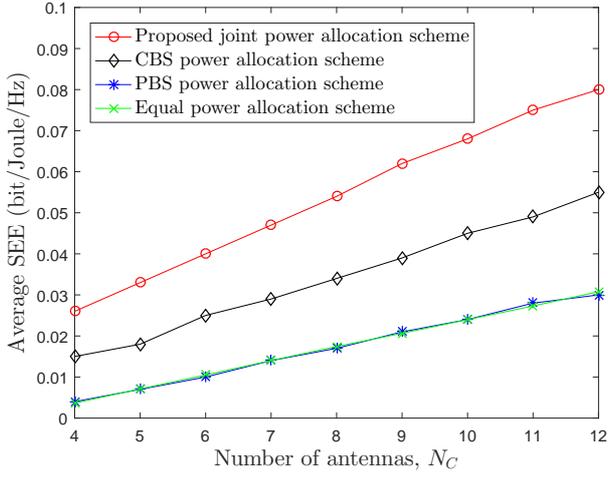}\\
  \caption{Average SEE versus the number of CBS's antennas, $N_C$, for different power allocation schemes with $I=8$, $N_P=4$, $N_E=3$, and the maximum transmit power of PBS and CBS, $P_{\textrm{PBS}}^{\textrm{total}}=30\textrm{dBm}$, $P_{\textrm{CBS}}^{\textrm{total}}=20\textrm{dBm}$.}\label{Fig6}}
\end{figure}
Fig. 6 depicts the average SEE results of the proposed joint power allocation scheme, CBS power allocation, PBS power allocation and equal power allocation schemes versus the number of CBS's antennas, $N_C$, in the cases of $I=8$, $N_P=4$, $N_E=3$, and the maximum transmit power of PBS and CBS, $P_{\textrm{PBS}}^{\textrm{total}}=30\textrm{dBm}$, $P_{\textrm{CBS}}^{\textrm{total}}=20\textrm{dBm}$. As shown in Fig. 5, the average SEE of the all schemes increases as $N_C$ increases, which means that the average SEE of OFDM-based CRNs can be further enhanced by employing more antennas of the CBS. Besides, the growth rate of proposed joint power allocation scheme is higher than the other power allocation schemes, showing that the number of antennas for joint optimal power allocation scheme has a more impact on the average SEE than the other power allocation schemes.

\begin{figure}[!h]
 \centering
 {\includegraphics[scale=0.6]{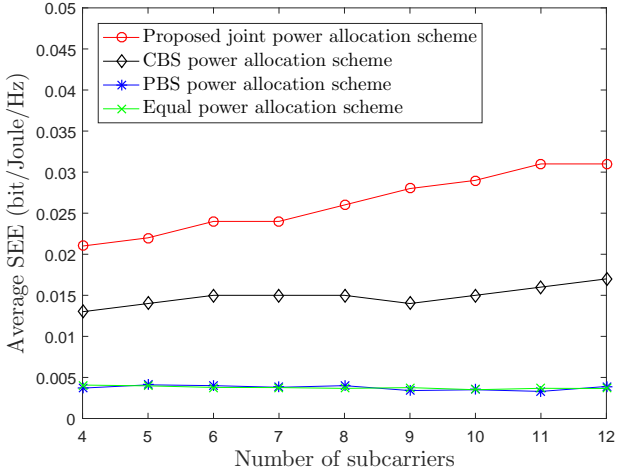}\\
  \caption{Average SEE versus the number of subcarriers, $I$, for different power allocation schemes with $N_P=N_C=4$, $N_E=3$, and the maximum transmit power of PBS and CBS, $P_{\textrm{PBS}}^{\textrm{total}}=30\textrm{dBm}$, $P_{\textrm{CBS}}^{\textrm{total}}=20\textrm{dBm}$.}\label{Fig7}}
\end{figure}
Fig. 7 shows the average SEE results of the proposed joint power allocation scheme versus the number of subcarriers, $I$, for different power allocation schemes with $N_P=N_C=4$, $N_E=3$, and the maximum transmit power of PBS and CBS, $P_{\textrm{PBS}}^{\textrm{total}}=30\textrm{dBm}$, $P_{\textrm{CBS}}^{\textrm{total}}=20\textrm{dBm}$. As observed, the proposed joint power allocation outperforms the other power allocation methods in terms of average SEE. Futhermore, giving the transmit power of PBS and CBS, as the number of subcarrier $I$ increases, the average SEE of the PBS power allocation and equal power allocation schemes almost keep unchanged. However, the average SEE of proposed joint power allocation and CBS power allocation approaches increase slightly. Besides, the proposed joint power allocation and CBS power allocation schemes obtain a higher average SEE than the PBS power allocation and equal power allocation methods, which indicates that the CBS transmit power allocation is more important than PBS power allocation in OFDM-based CRNs.

\section{Conclusion}

In this paper, we studied the power allocation of PBS and CBS across different OFDM subcarries in downlink OFDM-based CRNs. We first employed AN to improve the PLS of OFDM-based CRNs, and then formulated a power allocation problem to maximize the SEE based on instantaneous and statistical CSI of ED, where the circuit power consumption, minimum SR constraint, and minimum SR requirement were taken into consideration. New two-tier power allocation algorithms were presented to optimize the power allocation of PBS and CBS across different OFDM subcarriers. To be specific, with the help of the Dinkelbach's method and D.C. approaches, we converted the originally formulated non-convex problems into convex problems. Finally, numerical results showed that the proposed $\varepsilon$-optimal power allocation scheme obtains a higher SEE than conventional power allocation methods. Also, the proposed ICSI-SEEM and SCSI-SEEM schemes can improve the SEE of CRNs significantly compared with conventional SRM and EEM approaches.

\section*{Appendix A}
\section*{Proof of Theorem 1}
It is obvious that the problems (15) and (16) have the same feasible region ${\Re _1}$ for their same constraint conditions $C1$-$C4$. Firstly, we denote $( {{{{\bf{\mathord{\buildrel{\lower3pt\hbox{$\scriptscriptstyle\frown$}}
\over P} }}}_p},{{{\bf{\mathord{\buildrel{\lower3pt\hbox{$\scriptscriptstyle\frown$}}
\over P} }}}_s},{{{\bf{\mathord{\buildrel{\lower3pt\hbox{$\scriptscriptstyle\frown$}}
\over P} }}}_z}} ) \in {\Re _1}$ and $({\bf{\mathord{\buildrel{\lower3pt\hbox{$\scriptscriptstyle\frown$}}
\over P} }}_p^ * ,{\bf{\mathord{\buildrel{\lower3pt\hbox{$\scriptscriptstyle\frown$}}
\over P} }}_s^ * ,{\bf{\mathord{\buildrel{\lower3pt\hbox{$\scriptscriptstyle\frown$}}
\over P} }}_z^ * )\in {\Re _1}$ as the feasible and optimal solution of problem (15), respectively, so the maximum SEE $\eta _{\textrm{SEE}}^ * $ can be achieved by the following formula
\begin{equation}\tag{A.1}
\begin{split}
\eta _{{\rm{SEE}}}^ *  &= \!\!\mathop {\max }\limits_{{{\bf{P}}_p},{{\bf{P}}_s},{{\bf{P}}_z}} \frac{{\sum\limits_{i = 1}^I {\left[{{f_1}\left( {{P_{p,i}},{P_{s,i}},{P_{z,i}}} \right) - {f_2}\left( {{P_{p,i}},{P_{s,i}},{P_{z,i}}} \right)}\right ]} }}{{\sum\limits_{i = 1}^I {\left( {{P_{s,i}} + {P_{z,i}}} \right) + {P_b}} }}\\
& = \frac{{\sum\limits_{i = 1}^I {\left[ {{f_1}( {\mathord{\buildrel{\lower3pt\hbox{$\scriptscriptstyle\frown$}}
\over P} _{p,i}^ * ,\mathord{\buildrel{\lower3pt\hbox{$\scriptscriptstyle\frown$}}
\over P} _{s,i}^ * ,\mathord{\buildrel{\lower3pt\hbox{$\scriptscriptstyle\frown$}}
\over P} _{z,i}^ * } ) - {f_2}( {\mathord{\buildrel{\lower3pt\hbox{$\scriptscriptstyle\frown$}}
\over P} _{p,i}^ * ,\mathord{\buildrel{\lower3pt\hbox{$\scriptscriptstyle\frown$}}
\over P} _{s,i}^ * ,\mathord{\buildrel{\lower3pt\hbox{$\scriptscriptstyle\frown$}}
\over P} _{z,i}^ * } )}\right ]} }}{{\sum\limits_{i = 1}^I {( {\mathord{\buildrel{\lower3pt\hbox{$\scriptscriptstyle\frown$}}
\over P} _{s,i}^ *  + \mathord{\buildrel{\lower3pt\hbox{$\scriptscriptstyle\frown$}}
\over P} _{z,i}^ * }) + {P_b}} }}\\
&\ge \frac{{\sum\limits_{i = 1}^I {\left[ {{f_1}( {{{\mathord{\buildrel{\lower3pt\hbox{$\scriptscriptstyle\frown$}}
\over P} }_{p,i}},{{\mathord{\buildrel{\lower3pt\hbox{$\scriptscriptstyle\frown$}}
\over P} }_{s,i}},{{\mathord{\buildrel{\lower3pt\hbox{$\scriptscriptstyle\frown$}}
\over P} }_{z,i}}} ) - {f_2}( {{{\mathord{\buildrel{\lower3pt\hbox{$\scriptscriptstyle\frown$}}
\over P} }_{p,i}},{{\mathord{\buildrel{\lower3pt\hbox{$\scriptscriptstyle\frown$}}
\over P} }_{s,i}},{{\mathord{\buildrel{\lower3pt\hbox{$\scriptscriptstyle\frown$}}
\over P} }_{z,i}}} )} \right]} }}{{\sum\limits_{i = 1}^I {( {{{\mathord{\buildrel{\lower3pt\hbox{$\scriptscriptstyle\frown$}}
\over P} }_{s,i}} + {{\mathord{\buildrel{\lower3pt\hbox{$\scriptscriptstyle\frown$}}
\over P} }_{z,i}}} ) + {P_b}} }}.
\end{split}
\end{equation}
Based on the fact that $\sum\limits_{i = 1}^I {( {{{\mathord{\buildrel{\lower3pt\hbox{$\scriptscriptstyle\frown$}}
\over P} }_{s,i}} + {{\mathord{\buildrel{\lower3pt\hbox{$\scriptscriptstyle\frown$}}
\over P} }_{z,i}}})}  + {P_b} > 0$, (A.1) can be further transmitted into the following form
\begin{equation}\tag{A.2}
\begin{split}
\sum\limits_{i = 1}^I &{\left[ {{f_1}( {\mathord{\buildrel{\lower3pt\hbox{$\scriptscriptstyle\frown$}}
\over P} _{p,i}^ * ,\mathord{\buildrel{\lower3pt\hbox{$\scriptscriptstyle\frown$}}
\over P} _{s,i}^ * ,\mathord{\buildrel{\lower3pt\hbox{$\scriptscriptstyle\frown$}}
\over P} _{z,i}^ * } ) - {f_2}( {\mathord{\buildrel{\lower3pt\hbox{$\scriptscriptstyle\frown$}}
\over P} _{p,i}^ * ,\mathord{\buildrel{\lower3pt\hbox{$\scriptscriptstyle\frown$}}
\over P} _{s,i}^ * ,\mathord{\buildrel{\lower3pt\hbox{$\scriptscriptstyle\frown$}}
\over P} _{z,i}^ * } )} \right]} \\
& - \eta _{{\rm{SEE}}}^ * \left[ {\sum\limits_{i = 1}^I {( {\mathord{\buildrel{\lower3pt\hbox{$\scriptscriptstyle\frown$}}
\over P} _{s,i}^ *  + \mathord{\buildrel{\lower3pt\hbox{$\scriptscriptstyle\frown$}}
\over P} _{z,i}^ * } ) + {P_b}} } \right] = 0,
\end{split}
\end{equation}
\begin{equation}\tag{A.3}
\begin{split}
\sum\limits_{i = 1}^I &{\left[ {{f_1}( {{{\mathord{\buildrel{\lower3pt\hbox{$\scriptscriptstyle\frown$}}
\over P} }_{p,i}},{{\mathord{\buildrel{\lower3pt\hbox{$\scriptscriptstyle\frown$}}
\over P} }_{s,i}},{{\mathord{\buildrel{\lower3pt\hbox{$\scriptscriptstyle\frown$}}
\over P} }_{z,i}}} ) - {f_2}( {{{\mathord{\buildrel{\lower3pt\hbox{$\scriptscriptstyle\frown$}}
\over P} }_{p,i}},{{\mathord{\buildrel{\lower3pt\hbox{$\scriptscriptstyle\frown$}}
\over P} }_{s,i}},{{\mathord{\buildrel{\lower3pt\hbox{$\scriptscriptstyle\frown$}}
\over P} }_{z,i}}})} \right]}  \\
&- \eta _{{\rm{SEE}}}^ * \left[ {\sum\limits_{i = 1}^I {( {{{\mathord{\buildrel{\lower3pt\hbox{$\scriptscriptstyle\frown$}}
\over P} }_{s,i}} + {{\mathord{\buildrel{\lower3pt\hbox{$\scriptscriptstyle\frown$}}
\over P} }_{z,i}}} ) + {P_b}} } \right] \le 0.
\end{split}
\end{equation}
Combining (A.2) and (A.3), we can observe that the maximum value $f( {\eta _{\textrm{SEE}}^ * } )=0$ at the optimal solution $({\bf{\mathord{\buildrel{\lower3pt\hbox{$\scriptscriptstyle\frown$}}
\over P} }}_p^ * ,{\bf{\mathord{\buildrel{\lower3pt\hbox{$\scriptscriptstyle\frown$}}
\over P} }}_s^ * ,{\bf{\mathord{\buildrel{\lower3pt\hbox{$\scriptscriptstyle\frown$}}
\over P} }}_z^ * )$. Then, assuming $({\bf{\mathord{\buildrel{\lower3pt\hbox{$\scriptscriptstyle\smile$}}
\over P} }}_p^ * ,{\bf{\mathord{\buildrel{\lower3pt\hbox{$\scriptscriptstyle\smile$}}
\over P} }}_s^ * ,{\bf{\mathord{\buildrel{\lower3pt\hbox{$\scriptscriptstyle\smile$}}
\over P} }}_z^ * )\in {\Re _1}$ and $( {{{\bf{\mathord{\buildrel{\lower3pt\hbox{$\scriptscriptstyle\smile$}}
\over P} }}}_p},{{{\bf{\mathord{\buildrel{\lower3pt\hbox{$\scriptscriptstyle\smile$}}
\over P} }}}_s},{{{\bf{\mathord{\buildrel{\lower3pt\hbox{$\scriptscriptstyle\smile$}}
\over P} }}}_z} ) \in {\Re _1}$ are the optimal and feasible solution of problem (16), respectively, as well as $f( {\eta _{\textrm{SEE}}^ * } ) = 0$, that is
\begin{equation}\tag{A.4}
\begin{split}
f\left( {\eta _{{\rm{SEE}}}^ * } \right)&=\!\!\!\!\! \mathop {\max }\limits_{{{\bf{P}}_p},{{\bf{P}}_s},{{\bf{P}}_z}}\!\! \sum\limits_{i = 1}^I \left[ {{f_1}\!\left( {{P_{p,i}},{P_{s,i}},{P_{z,i}}} \right)\! -\! {f_2}\!\left( {{P_{p,i}},{P_{s,i}},{P_{z,i}}} \right)} \right]\\
&\qquad  - {\eta _{\textrm{SEE}}^ *}\left[ {\sum\limits_{i = 1}^I {\left( {{P_{s,i}} + {P_{z,i}}} \right) + {P_b}} } \right]\\
 & = \sum\limits_{i = 1}^I {\left[ {{f_1}( {\mathord{\buildrel{\lower3pt\hbox{$\scriptscriptstyle\smile$}}
\over P} _{p,i}^ * ,\mathord{\buildrel{\lower3pt\hbox{$\scriptscriptstyle\smile$}}
\over P} _{s,i}^ * ,\mathord{\buildrel{\lower3pt\hbox{$\scriptscriptstyle\smile$}}
\over P} _{z,i}^ * } ) - {f_2}( {\mathord{\buildrel{\lower3pt\hbox{$\scriptscriptstyle\smile$}}
\over P} _{p,i}^ * ,\mathord{\buildrel{\lower3pt\hbox{$\scriptscriptstyle\smile$}}
\over P} _{s,i}^ * ,\mathord{\buildrel{\lower3pt\hbox{$\scriptscriptstyle\smile$}}
\over P} _{z,i}^ * } )} \right]} \\
&\qquad-\eta _{\textrm{SEE}}^ * \left[ {\sum\limits_{i = 1}^I {( {\mathord{\buildrel{\lower3pt\hbox{$\scriptscriptstyle\smile$}}
\over P} _{s,i}^ *  + \mathord{\buildrel{\lower3pt\hbox{$\scriptscriptstyle\smile$}}
\over P} _{z,i}^ * } ) + {P_b}} } \right]\\
&=0\\
& \ge \sum\limits_{i = 1}^I {\left[ {{f_1}( {{{\mathord{\buildrel{\lower3pt\hbox{$\scriptscriptstyle\smile$}}
\over P} }_{p,i}},{{\mathord{\buildrel{\lower3pt\hbox{$\scriptscriptstyle\smile$}}
\over P} }_{s,i}},{{\mathord{\buildrel{\lower3pt\hbox{$\scriptscriptstyle\smile$}}
\over P} }_{z,i}}} ) - {f_2}( {{{\mathord{\buildrel{\lower3pt\hbox{$\scriptscriptstyle\smile$}}
\over P} }_{p,i}},{{\mathord{\buildrel{\lower3pt\hbox{$\scriptscriptstyle\smile$}}
\over P} }_{s,i}},{{\mathord{\buildrel{\lower3pt\hbox{$\scriptscriptstyle\smile$}}
\over P} }_{z,i}}})} \right]} \\
&\qquad- \eta _{\textrm{SEE}}^ * \left[ {\sum\limits_{i = 1}^I {( {{{\mathord{\buildrel{\lower3pt\hbox{$\scriptscriptstyle\smile$}}
\over P} }_{s,i}} + {{\mathord{\buildrel{\lower3pt\hbox{$\scriptscriptstyle\smile$}}
\over P} }_{z,i}}} ) + {P_b}} } \right].
\end{split}
\end{equation}
After some operations, we can achieve the following fractional formula
\begin{equation}\tag{A.5}
\begin{split}
&\frac{{\sum\limits_{i = 1}^I {\left[ {{f_1}({{\mathord{\buildrel{\lower3pt\hbox{$\scriptscriptstyle\smile$}}
\over P} }_{p,i}},{{\mathord{\buildrel{\lower3pt\hbox{$\scriptscriptstyle\smile$}}
\over P} }_{s,i}},{{\mathord{\buildrel{\lower3pt\hbox{$\scriptscriptstyle\smile$}}
\over P} }_{z,i}}) - {f_2}({{\mathord{\buildrel{\lower3pt\hbox{$\scriptscriptstyle\smile$}}
\over P} }_{p,i}},{{\mathord{\buildrel{\lower3pt\hbox{$\scriptscriptstyle\smile$}}
\over P} }_{s,i}},{{\mathord{\buildrel{\lower3pt\hbox{$\scriptscriptstyle\smile$}}
\over P} }_{z,i}})} \right]} }}{{\sum\limits_{i = 1}^I {({{\mathord{\buildrel{\lower3pt\hbox{$\scriptscriptstyle\smile$}}
\over P} }_{s,i}} + {{\mathord{\buildrel{\lower3pt\hbox{$\scriptscriptstyle\smile$}}
\over P} }_{z,i}}) + {P_b}} }}\\
\le& \eta _{\textrm{SEE}}^ * \\
 =& \frac{{\sum\limits_{i = 1}^I {\left[ {{f_1}(\mathord{\buildrel{\lower3pt\hbox{$\scriptscriptstyle\smile$}}
\over P} _{p,i}^ * ,\mathord{\buildrel{\lower3pt\hbox{$\scriptscriptstyle\smile$}}
\over P} _{s,i}^ * ,\mathord{\buildrel{\lower3pt\hbox{$\scriptscriptstyle\smile$}}
\over P} _{z,i}^ * ) - {f_2}(\mathord{\buildrel{\lower3pt\hbox{$\scriptscriptstyle\smile$}}
\over P} _{p,i}^ * ,\mathord{\buildrel{\lower3pt\hbox{$\scriptscriptstyle\smile$}}
\over P} _{s,i}^ * ,\mathord{\buildrel{\lower3pt\hbox{$\scriptscriptstyle\smile$}}
\over P} _{z,i}^ * )} \right]} }}{{\sum\limits_{i = 1}^I {(\mathord{\buildrel{\lower3pt\hbox{$\scriptscriptstyle\smile$}}
\over P} _{s,i}^ *  + \mathord{\buildrel{\lower3pt\hbox{$\scriptscriptstyle\smile$}}
\over P} _{z,i}^ * ) + {P_b}} }}.
\end{split}
\end{equation}
From (A.5), it is easy to find that $(  {\bf{\mathord{\buildrel{\lower3pt\hbox{$\scriptscriptstyle\smile$}}
\over P} }}_p^ * ,{\bf{\mathord{\buildrel{\lower3pt\hbox{$\scriptscriptstyle\smile$}}
\over P} }}_s^ * ,{\bf{\mathord{\buildrel{\lower3pt\hbox{$\scriptscriptstyle\smile$}}
\over P} }}_z^ *)$ is also the optimal solution of (15). Therefore, we can obtain that $({\bf{\mathord{\buildrel{\lower3pt\hbox{$\scriptscriptstyle\frown$}}
\over P} }}_p^ * ,{\bf{\mathord{\buildrel{\lower3pt\hbox{$\scriptscriptstyle\frown$}}
\over P} }}_s^ * ,{\bf{\mathord{\buildrel{\lower3pt\hbox{$\scriptscriptstyle\frown$}}
\over P} }}_z^ * )$ is equal to $( {\bf{\mathord{\buildrel{\lower3pt\hbox{$\scriptscriptstyle\smile$}}
\over P} }}_p^ * ,{\bf{\mathord{\buildrel{\lower3pt\hbox{$\scriptscriptstyle\smile$}}
\over P} }}_s^ * ,{\bf{\mathord{\buildrel{\lower3pt\hbox{$\scriptscriptstyle\smile$}}
\over P} }}_z^ * )$ if and only if $f\left( {\eta _{\textrm{\textrm{SEE}}}^ * } \right) = 0$.

\section*{Appendix B}
\section*{Proof of the Convergence}
Assuming that $( {\vec{\bf {P}}}_p^{n+1},{\vec{\bf {P}}}_s^{n+1},{\vec{\bf {P}}}_z^{n+1} )$ and $( {\vec{\bf {P}}}_p^{n},{\vec{\bf {P}}}_s^{n},{\vec{\bf {P}}}_z^{n} )$ are feasible solutions of (22) at iterations $n+1$ and $n$, respectively, and using (19) and (20), we can obtain
\begin{equation}\tag{B.1}
\begin{split}
&\quad{f_2}( {\vec P_{p,i}^{n + 1},\vec P_{s,i}^{n + 1},\vec P_{z,i}^{n + 1}})\!\le\! {f_2}( {\vec P_{p,i}^n,\vec P_{s,i}^n,\vec P_{z,i}^n})\\
&\! +\! \frac{{{b_i}( {\vec P_{p,i}^{n + 1} \!\!-\! \vec P_{p,i}^n})}}{{( {{b_i}\vec P_{p,i}^n \!+\! \sigma _{c,i}^2} )\!\ln 2}}\!+\! \frac{{{\rm{Tr}}\!\!\left[ {{{\bf{c}}_i}{{( \vec {\bf{\Omega }} _i^n )}^{ - 1}}\!( {\vec P_{p,i}^{n + 1} \!-\! \vec P_{p,i}^n} )} \right]}}{{\ln 2}}\\
&\!+\! \frac{{{\rm{Tr}}\!\!\left[ {{{\bf{f}}_i}{{( \vec {\bf{\Omega }} _i^n )}^{ - 1}}\!( {\vec P_{s,i}^{n + 1}\! -\! \vec P_{s,i}^n} )} \right]}}{{\ln 2}} \!+\!{\frac{{{\rm{Tr}}\!\!\left[ {{{\bf{g}}_i}{{( \vec {\bf{\Omega }} _i^n )}^{ - 1}}\!( {\vec P_{z,i}^{n + 1} \!-\! \vec P_{z,i}^n} )} \right]}}{{\ln 2}}},
\end{split}
\end{equation}
and
\begin{equation}\tag{B.2}
\begin{split}
&\quad{g_2}( {\vec P_{p,i}^{n + 1},\vec P_{s,i}^{n + 1},\vec P_{z,i}^{n + 1}} ) \!\le\! {g_2}( {\vec P_{p,i}^n,\vec P_{s,i}^n,\vec P_{z,i}^n})\\
&\! +\! \frac{{{d_i}( {\vec P_{s,i}^{n + 1}\! -\! \vec P_{s,i}^n})}}{{( {{d_i}\vec P_{s,i}^n \!+\! \sigma _{p,i}^2} )\!\ln 2}}\!+\! \frac{{{\rm{Tr}}\!\!\left[ {{{\bf{c}}_i}{{( \vec {\bf{\Omega }} _i^n )}^{ - 1}}\!( {\vec P_{p,i}^{n + 1}\! -\! \vec P_{p,i}^n} )} \right]}}{{\ln 2}} \\
&\! +\! \frac{{{\rm{Tr}}\!\!\left[ {{{\bf{f}}_i}{{( \vec {\bf{\Omega }} _i^n )}^{ - 1}}\!( {\vec P_{s,i}^{n + 1}\! -\! \vec P_{s,i}^n} )} \right]}}{{\ln 2}} \! +\!{\frac{{{\rm{Tr}}\!\!\left[ {{{\bf{g}}_i}{{( \vec {\bf{\Omega }} _i^n )}^{ - 1}}\!( {\vec P_{z,i}^{n + 1} \!-\! \vec P_{z,i}^n} )} \right]}}{{\ln 2}}},
\end{split}
\end{equation}
where $\vec {\bf{\Omega }} _i^n = {{{\bf{c}}_i}\vec P_{p,i}^n\! +\! {{\bf{f}}_i}\vec P_{s,i}^n\! + \!{{\bf{g}}_i}\vec P_{z,i}^n \!+\! \sigma _{e,i}^2{{\bf{I}}_{{N_E}}}} $. Substituting feasible solutions of (22) into $C1$ and $C2$ of (16), we can obtain
\begin{equation}\tag{B.3}
\begin{split}
&\quad {f_1}(\vec P_{p,i}^{n + 1},\vec P_{s,i}^{n + 1},\vec P_{z,i}^{n + 1}) - {f_2}(\vec P_{p,i}^{n + 1},\vec P_{s,i}^{n + 1},\vec P_{z,i}^{n + 1})   \\
&\ge {f_1}(\vec P_{p,i}^{n + 1},\vec P_{s,i}^{n + 1},\vec P_{z,i}^{n + 1}) - {f_2}(\vec P_{p,i}^n,\vec P_{s,i}^n,\vec P_{z,i}^n)\\
&\! -\! \frac{{{\rm{Tr}}\!\!\left[ {{{\bf{c}}_i}{{(\vec {\bf{\Omega }} _i^n)}^{ - 1}}\!(\vec P_{p,i}^{n + 1}\!\! -\! \vec P_{p,i}^n)} \right]}}{{\ln 2}} \! -\! \frac{{{\rm{Tr}}\!\!\left[ {{{\bf{f}}_i}{{(\vec {\bf{\Omega }} _i^n)}^{ - 1}}\!(\vec P_{s,i}^{n + 1} \!\!-\! \vec P_{s,i}^n)} \right]}}{{\ln 2}}\\
& \!-\! \frac{{{\rm{Tr}}\!\!\left[ {{{\bf{g}}_i}{{(\vec {\bf{\Omega }} _i^n)}^{ - 1}}\!(\vec P_{z,i}^{n + 1} \!\!-\! \vec P_{z,i}^n)} \right]}}{{\ln 2}}\! -\! \frac{{{b_i}(\vec P_{p,i}^{n + 1}\!\! -\! \vec P_{p,i}^n)}}{{({b_i}\vec P_{p,i}^n\! +\! \sigma _{c,i}^2)\ln 2}} \!\!\ge\!\! R_{CU}^{\min },{\kern 1pt}\forall i,
\end{split}
\end{equation}
and
\begin{equation}\tag{B.4}
\begin{split}
&\quad{g_1}(\vec P_{p,i}^{n + 1},\vec P_{s,i}^{n + 1},\vec P_{z,i}^{n + 1}) - {g_2}(\vec P_{p,i}^{n + 1},\vec P_{s,i}^{n + 1},\vec P_{z,i}^{n + 1}) \\
& \ge {g_1}(\vec P_{p,i}^{n + 1},\vec P_{s,i}^{n + 1},\vec P_{z,i}^{n + 1}) - {g_2}(\vec P_{p,i}^n,\vec P_{s,i}^n,\vec P_{z,i}^n)\\
&\! -\!\frac{{{\rm{Tr}}\!\!\left[ {{{\bf{c}}_i}{{(\vec {\bf{\Omega }} _i^n)}^{ - 1}}\!(\vec P_{p,i}^{n + 1}\!\! -\! \vec P_{p,i}^n)} \right]}}{{\ln 2}}\!-\!\frac{{{\rm{Tr}}\!\!\left[ {{{\bf{f}}_i}{{(\vec {\bf{\Omega }} _i^n)}^{ - 1}}\!(\vec P_{s,i}^{n + 1} \!\!-\! \vec P_{s,i}^n)} \right]}}{{\ln 2}}\\
&\!-\!\frac{{{\rm{Tr}}\!\!\left[ {{{\bf{g}}_i}{{(\vec {\bf{\Omega }} _i^n)}^{ - 1}}\!(\vec P_{z,i}^{n + 1} \!-\! \vec P_{z,i}^n)} \right]}}{{\ln 2}} \!-\!\frac{{{d_i}(\vec P_{s,i}^{n + 1}\!\! -\! \vec P_{s,i}^n)}}{{({d_i}\vec P_{s,i}^n\! +\! \sigma _{p,i}^2)\ln 2}}\!\! \ge\!\! R_{PU}^{\min },{\kern 1pt}\forall i.
\end{split}
\end{equation}
From (B.3) and (B.4), we can observe that the feasible solutions of (22) are also suitable for (16).

According to (19), we also obtain
\begin{equation}\tag{B.5}
\begin{split}
&\quad{f_2}(\bar P_{p,i}^{n + 1},\bar P_{s,i}^{n + 1},\bar P_{z,i}^{n + 1})\!\le\! {f_2}(\bar P_{p,i}^n,\bar P_{s,i}^n,\bar P_{z,i}^n)\\
&\! + \!\frac{{{b_i}(\bar P_{p,i}^{n + 1}\! -\! \bar P_{p,i}^n)}}{{({b_i}\bar P_{p,i}^n \!+\! \sigma _{c,i}^2)\!\ln 2}}\!+\! \frac{{{\rm{Tr}}\!\!\left[ {{{\bf{c}}_i}{{({\bf{\bar \Omega }}_i^n)}^{ - 1}}\!(\bar P_{p,i}^{n + 1}\! -\! \bar P_{p,i}^n)} \right]}}{{\ln 2}}\\
&\!+ \! \frac{{{\rm{Tr}}\!\!\left[ {{{\bf{f}}_i}{{({\bf{\bar \Omega }}_i^n)}^{ - 1}}\!(\bar P_{s,i}^{n + 1}\! -\! \bar P_{s,i}^n)} \right]}}{{\ln 2}}\! +\! \frac{{{\rm{Tr}}\!\!\left[ {{{\bf{g}}_i}{{({\bf{\bar \Omega }}_i^n)}^{ - 1}}\!(\bar P_{z,i}^{n + 1} \!-\! \bar P_{z,i}^n)} \right]}}{{\ln 2}}.
\end{split}
\end{equation}
Then, following the iterative procedure in (22), we arrive at
\begin{equation}\tag{B.6}
\begin{split}
&\sum\limits_{i = 1}^I {\left\{ {{f_1}(\bar P_{p,i}^{n + 1},\bar P_{s,i}^{n + 1},\bar P_{z,i}^{n + 1})} \right.}  - {f_2}(\bar P_{p,i}^n,\bar P_{s,i}^n,\bar P_{z,i}^n)\\
& \!-\! \frac{{{b_i}(\bar P_{p,i}^{n + 1} - \bar P_{p,i}^n)}}{{({b_i}\bar P_{p,i}^n + \sigma _{c,i}^2)ln2}} - \frac{{{\rm{Tr}}\!\!\left[ {{{\bf{c}}_i}{{({\bf{\bar \Omega }}_i^n)}^{ - 1}}\!(\bar P_{p,i}^{n + 1}\!\! - \!\bar P_{p,i}^n)} \right]}}{{\ln 2}}\\
& \!-\! \left. {\frac{{{\rm{Tr}}\!\!\left[ {{{\bf{f}}_i}{{({\bf{\bar \Omega }}_i^n)}^{ - 1}}\!(\bar P_{s,i}^{n + 1}\!\! -\! \bar P_{s,i}^n)} \right]}}{{\ln 2}}\! -\! \frac{{{\rm{Tr}}\!\!\left[ {{{\bf{g}}_i}{{({\bf{\bar \Omega }}_i^n)}^{ - 1}}\!(\bar P_{z,i}^{n + 1} \!\!-\! \bar P_{z,i}^n)} \right]}}{{\ln 2}}} \!\right\}\\
& \!-\! {\eta _{{\rm{SEE}}}}\left[ {\sum\limits_{i = 1}^I {(\bar P_{s,i}^{n + 1} + \bar P_{z,i}^{n + 1}) + {P_b}} } \right]\\
&\! =\! \mathop {\max }\limits_{{{\bf{P}}_p},{{\bf{P}}_s},{{\bf{P}}_z}} \sum\limits_{i = 1}^I {\left\{ {{f_1}} \right.({P_{p,i}},{P_{s,i}},{P_{z,i}}) - {f_2}(\bar P_{p,i}^n,\bar P_{s,i}^n,\bar P_{z,i}^n)} \\
& \!-\! \frac{{{b_i}({P_{p,i}} - \bar P_{p,i}^n)}}{{({b_i}\bar P_{p,i}^n + \sigma _{c,i}^2){\ln2}}} - \frac{{{\rm{Tr}}\!\!\left[ {{{\bf{c}}_i}{{({\bf{\bar \Omega }}_i^n)}^{ - 1}}\!({P_{p,i}}\! -\! \bar P_{p,i}^n)} \right]}}{{\ln 2}}\\
& \!-\! \left. {\frac{{{\rm{Tr}}\left[ {{{\bf{f}}_i}{{({\bf{\bar \Omega }}_i^n)}^{ - 1}}\!({P_{s,i}} \!-\! \bar P_{s,i}^n)} \right]}}{{\ln 2}} \!-\! \frac{{{\rm{Tr}}\!\!\left[ {{{\bf{g}}_i}{{({\bf{\bar \Omega }}_i^n)}^{ - 1}}\!({P_{z,i}}\! -\! \bar P_{z,i}^n)} \right]}}{{\ln 2}}}\! \right\}\\
& - {\eta _{{\rm{SEE}}}}\left[ {\sum\limits_{i = 1}^I {({P_{s,i}} + {P_{z,i}}) + {P_b}} } \right]\\
&\ge \sum\limits_{i = 1}^I {\left[ {{f_1}(\bar P_{p,i}^n,\bar P_{s,i}^n,\bar P_{z,i}^n) - {f_2}(\bar P_{p,i}^n,\bar P_{s,i}^n,\bar P_{z,i}^n)} \right]} \\
& \quad- {\eta _{{\rm{SEE}}}}\left[ {\sum\limits_{i = 1}^I {(\bar P_{s,i}^n + \bar P_{z,i}^n) + {P_b}} } \right].
\end{split}
\end{equation}
Substituting (B.5) into (B.6), we can further have
\begin{equation}\tag{B.7}
\begin{split}
&\sum\limits_{i = 1}^I {\left[ {{f_1}(\bar P_{p,i}^{n + 1},\bar P_{s,i}^{n + 1},\bar P_{z,i}^{n + 1}) - {f_2}(\bar P_{p,i}^{n + 1},\bar P_{s,i}^{n + 1},\bar P_{z,i}^{n + 1})} \right]} \\
& \quad- {\eta _{{\rm{SEE}}}}\left[ {\sum\limits_{i = 1}^I {(\bar P_{s,i}^{n + 1} + \bar P_{z,i}^{n + 1}) + {P_b}} } \right]\\
&\ge \sum\limits_{i = 1}^I {\left\{ {{f_1}(\bar P_{p,i}^{n + 1},\bar P_{s,i}^{n + 1},\bar P_{z,i}^{n + 1})} \right.}  - {f_2}(\bar P_{p,i}^n,\bar P_{s,i}^n,\bar P_{z,i}^n)\\
& \!-\! \frac{{{b_i}(\bar P_{p,i}^{n + 1} - \bar P_{p,i}^n)}}{{({b_i}\bar P_{p,i}^n + \sigma _{c,i}^2){\ln2}}} - \frac{{{\rm{Tr}}\!\!\left[ {{{\bf{c}}_i}{{({\bf{\bar \Omega }}_i^n)}^{ - 1}}\!(\bar P_{p,i}^{n + 1}\!\! - \!\bar P_{p,i}^n)} \right]}}{{\ln 2}}\\
& \!-\! \left. {\frac{{{\rm{Tr}}\!\!\left[ {{{\bf{f}}_i}{{({\bf{\bar \Omega }}_i^n)}^{ - 1}}\!(\bar P_{s,i}^{n + 1}\!\! -\! \bar P_{s,i}^n)} \right]}}{{\ln 2}}\! -\! \frac{{{\rm{Tr}}\!\!\left[ {{{\bf{g}}_i}{{({\bf{\bar \Omega }}_i^n)}^{ - 1}}\!(\bar P_{z,i}^{n + 1} \!\!-\! \bar P_{z,i}^n)} \right]}}{{\ln 2}}} \!\right\}\\
& \!-\! {\eta _{{\rm{SEE}}}}\left[ {\sum\limits_{i = 1}^I {(\bar P_{s,i}^{n + 1} + \bar P_{z,i}^{n + 1}) + {P_b}} } \right]\\
& \ge \sum\limits_{i = 1}^I {\left[ {{f_1}(\bar P_{p,i}^n,\bar P_{s,i}^n,\bar P_{z,i}^n) - {f_2}(\bar P_{p,i}^n,\bar P_{s,i}^n,\bar P_{z,i}^n)} \right]} \\
& \quad- {\eta _{{\rm{SEE}}}}\left[ {\sum\limits_{i = 1}^I {(\bar P_{s,i}^n + \bar P_{z,i}^n) + {P_b}} } \right].
\end{split}
\end{equation}
From (B.7), we can observe that the proposed iterative procedure is monotonically non-decreasing with the increasing of iterative numbers. In addition, by employing the transmit power constraints of PBS and CBS, i.e., $\sum\limits_{i = 1}^I {{P_{p,i}}}  \le P_{{\rm{PBS}}}^{{\rm{total}}}$ and $\sum\limits_{i = 1}^I {\left( {{P_{s,i}} + {P_{z,i}}} \right)}  \le P_{{\rm{CBS}}}^{{\rm{total}}}$, the upper bound of the objective function can be given by
\begin{equation}\tag{B.8}
\begin{split}
&\quad \sum\limits_{i = 1}^I {\left[ {{f_1}\left( {{P_{p,i}},{P_{s,i}},{P_{z,i}}} \right) - {f_2}\left( {{P_{p,i}},{P_{s,i}},{P_{z,i}}} \right)} \right]}\\
&-{\eta _{\textrm{SEE}}}\!\!\left[ {\sum\limits_{i = 1}^I {\left( {{P_{s,i}} \!+ \!{P_{z,i}}} \right) \!+ \!{P_b}} } \right]\!\le\!\sum\limits_{i = 1}^I {\left[ {{{\log }_2}( {1\! +\! \frac{{{e_i}{P_{s,i}}}}{{{b_i}{P_{p,i}}\! +\! \sigma _{c,i}^2}}} )} \right]} \\
 &\le \frac{{\max ({e_i})P_{\textrm{CBS}}^{\textrm{total}}}}{{\Delta f{N_0}\ln 2}}.
\end{split}
\end{equation}
Combining (B.7) and (B.8), we can guarantee that the iterative procedure in (22) will converge to an $\varepsilon$-optimal solution of (16) after sufficient iterations.



\begin{IEEEbiography}[{\includegraphics[width=1in,height=1.25in]{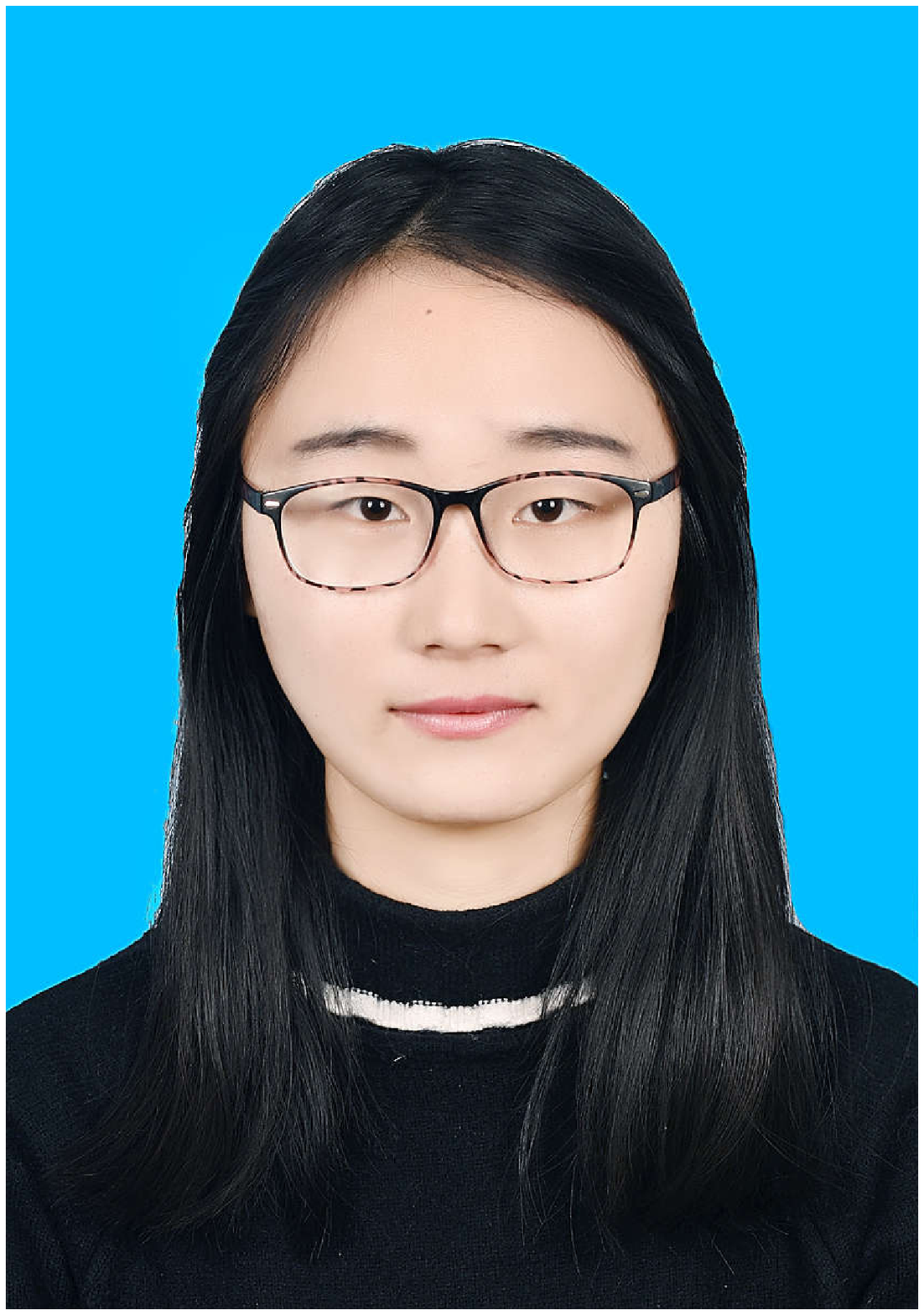}}]{Yuhan Jiang}
received the B.Eng. degree in Communication Engineering from Nantong University, Nantong, China, in July 2016. She is currently pursuing the Ph.D. degree in Signal and Information Processing at the Nanjing University of Posts and Telecommunications. Her research interests include cognitive radio, physical-layer security and green communications.
\end{IEEEbiography}

\begin{IEEEbiography}[{\includegraphics[width=1in,height=1.25in]{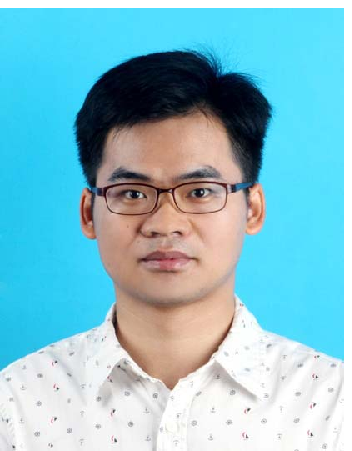}}]{Yulong Zou} (SM'13) is a Full Professor and Doctoral Supervisor at the Nanjing University of Posts and Telecommunications (NUPT), Nanjing, China. He received the B.Eng. degree in information engineering from NUPT, Nanjing, China, in July 2006, the first Ph.D. degree in electrical engineering from the Stevens Institute of Technology, New Jersey, USA, in May 2012, and the second Ph.D. degree in signal and information processing from NUPT, Nanjing, China, in July 2012.

Dr. Zou was awarded the 9th IEEE Communications Society Asia-Pacific Best Young Researcher in 2014. He has served as an editor for the IEEE Communications Surveys \& Tutorials, IEEE Communications Letters, EURASIP Journal on Advances in Signal Processing, IET Communications, and China Communications. In addition, he has acted as TPC members for various IEEE sponsored conferences, e.g., IEEE ICC/GLOBECOM/WCNC/VTC/ICCC, etc.
\end{IEEEbiography}

\begin{IEEEbiography}[{\includegraphics[width=1in,height=1.25in]{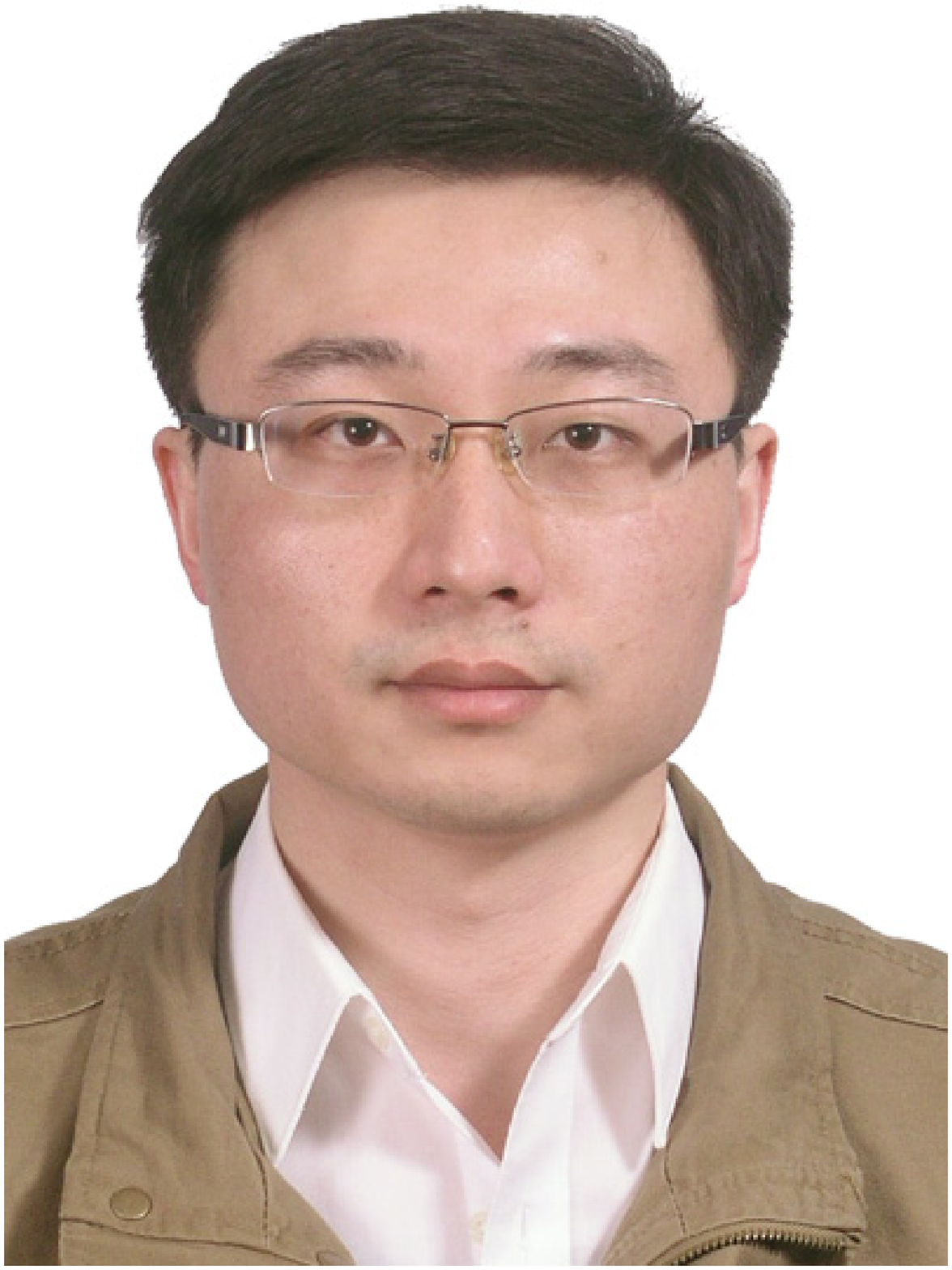}}]{Jian Ouyang} (M'5) received the B.S., M.S. and Ph.D. degrees from Nanjing University of Aeronautics and Astronautics, Nanjing, China, in 2007, 2010 and 2014, respectively. Since July 2014, he has been a full-time faculty member with the College of Telecommunications and Information Engineering, Nanjing University of Posts and Telecommunications, Nanjing, China. From 2015 to 2016, he was a Postdoctoral Fellow with the Department of Electrical and Computer Engineering, Concordia University, Montreal, Canada. His research interests include cooperative and relay communications, physical layer security and green communications.
\end{IEEEbiography}

\begin{IEEEbiography}[{\includegraphics[width=1in,height=1.25in]{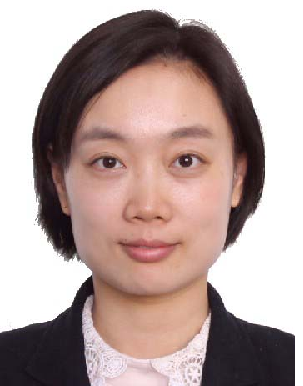}}]{Jia Zhu} is an Associate Professor at the Nanjing University of Posts and Telecommunications (NUPT), Nanjing, China. She received the B.Eng. degree in Computer Science and Technology from the Hohai University, Nanjing, China, in July 2005, and the Ph.D. degree in Signal and Information Processing from the Nanjing University of Posts and Telecommunications, Nanjing, China, in April 2010. From June 2010 to June 2012, she was a Postdoctoral Research Fellow at the Stevens Institute of Technology, New Jersey, the United States. Since November 2012, she has been a full-time faculty member with the Telecommunication and Information School of NUPT, Nanjing, China. Her general research interests include the cognitive radio, physical-layer security and communications theory.\end{IEEEbiography}

\clearpage


\begin{thebibliography}{11}

\bibitem{IEEEhowto:1}
C. Fan, B. Li, C. Zhao, and A. Nallanathan, ``Two-dimensional distributed spectrum reusing in cognitive radio network: Based on game theory," in \emph{Proc. IEEE Conf. Commun. (ICC)}, Paris, 2017, pp. 1-6.

\bibitem{IEEEhowto:2}
J. Zhu and Y. Zou, ``Cognitive network cooperation for green cellular networks," \emph{IEEE Access}, vol. 4, pp. 849-857, Feb. 2016.

\bibitem{IEEEhowto:3}
K. An, M. Lin, J. Ouyang, and W.-P. Zhu, ``Secure transmission in cognitive satellite terrestrial networks," \emph{IEEE J. Sel. Areas Commun.}, vol. 34, no. 11, pp. 3025-3037, Nov. 2016.


\bibitem{IEEEhowto:4}
D. B. Rawat, ``Evaluating performance of cognitive radio users in MIMO-OFDM-based wireless networks," \emph{IEEE Wireless Commun. Lett.}, vol. 5, no. 5, pp. 476-479, Oct. 2016.

\bibitem{IEEEhowto:5}
Y. Zou, J. Zhu, X. Wang, and L. Hanzo, ``A survey on wireless security: technical challenges, recent advances, and future trends," \emph{Proc. of the IEEE}, vol. 104, no. 9, pp. 1727-1765, Sept. 2016.


\bibitem{IEEEhowto:6}
Y. Zou, J. Zhu, L. Yang, Y.-C. Liang, and Y.-D. Yao, ``Securing physical-layer communications for cognitive radio networks," \emph{IEEE Commun. Mag.}, vol. 53, no. 9, pp. 48-54, Sept. 2015.

\bibitem{IEEEhowto:7}
M. Lin, J. Ouyang, and W.-P. Zhu, ``Joint beamforming and power control for device-to-device communications underlaying cellular networks," \emph{IEEE J. Sel. Areas Commun.}, vol. 34, no. 1, pp. 138-150, Jan. 2016.


\bibitem{IEEEhowto:8}
Y. Zhang, Z. Yang, A. Liu, and Y. Zou, ``Secure transmission over the wiretap channel using polar codes and artificial noise," \emph{IET Commun.}, vol. 11, pp. 377-384, 2017.

\bibitem{IEEEhowto:9}
T. M. Hoang, T. Q. Duong, N.-S. Vo, and C. Kundu, ``Physical layer security in cooperative energy harvesting networks with a friendly jammer," \emph{IEEE wireless Commun. Lett.}, vol. 6, no. 2, pp. 174-177, Apr. 2017.

\bibitem{IEEEhowto:10}
H. Wu, X. Tao, Z. Han, N. Li, and J. Xu, ``Secure transmission in MISOME wiretap channel with multiple assisting jammers: maximum secrecy rate and optimal power allocation," \emph{IEEE Trans. Commun.}, vol. 65, no. 2, pp. 775-789, Feb. 2017.

\bibitem{IEEEhowto:11}
J. P. Vilela, M. Bloch, J. Barros, and S. W. McLaughlin, ``Wireless secrecy regions with friendly jamming." \emph{IEEE Trans. Inf. Forensics Security}, vol. 6, no. 2, pp. 256-266, Jun. 2011.


\bibitem{IEEEhowto:12}
H. Xu, B. Zhu, J. Liu, and A. Zhou, ``Robust beamforming design for secure multiuser MISO interference channel," \emph{IEEE Commun. Lett.}, vol. 21, no. 4, pp. 833-836, Apr. 2017.

\bibitem{IEEEhowto:13}
V.-D. Nguyen, T. Q. Duong, O. A. Dobre, and O.-S. Shin, ``Joint information and jamming beamforming for secrecy rate maximization in cognitive radio networks," \emph{IEEE Trans. Inf. Forensics Security}, vol. 11, no. 11, pp. 2609-2623, Nov. 2016.

\bibitem{IEEEhowto:14}
J. Zhu, R. Schober, and V. K. Bhargava, ``Linear precoding of data and artificial noise in secure massive MIMO systems," \emph{IEEE Trans. Wireless Commun.}, vol. 15, no. 3, pp. 2245-2261, Mar. 2016.

\bibitem{IEEEhowto:15}
S. Yan, N. Yang, Ingmar Land, R. Malaney, and J. Yuan, ``Three artificial-noise-aided secure transmission schemes in wiretap channels," \emph{IEEE Trans. Veh. Technol.}, to appear, Dec. 2017.

\bibitem{IEEEhowto:16}
S. Yan, X. Zhou, N. Yang, B. He, and T. D. Abhayapala, ``Artificial-noise-aided secure transmission in wiretap channels with transmitter-side correlation," \emph{IEEE Trans. Wireless Commun.}, vol. 15, no. 12, pp. 8286-8297, Dec. 2016.

\bibitem{IEEEhowto:17}
N. Yang, S. Yan, J. Yuan, R. Malaney, R. Subramanian, and I. Land, ``Artificial noise: transmission optimization in Multi-Input Single-Output wiretap channels," \emph{IEEE Trans. Commun.}, vol. 63, no. 5, pp. 1771-1783, May 2015.

\bibitem{IEEEhowto:18}
H. Shokri-Ghadikolaei, I. Glaropoulos, V. Fodor, C. Fischione, and A. Ephremides, ``Green sensing and access: energy-throughput trade-offs in cognitive networking," \emph{IEEE Commun. Mag.}, vol. 53, no. 11, pp. 199-207, Nov. 2015.

\bibitem{IEEEhowto:19}
W. R. Mili, L. Musavian, K. A. Hamdi, and F. Marvasti, ``How to increase energy efficiency in cognitive radio networks," \emph{IEEE Trans. Commun.}, vol. 64, no. 5, pp.1829-1843, May 2016.

\bibitem{IEEEhowto:20}
F. Gabry, A. Zappone, R. Thobaben, \emph{et al.,} ``Energy efficient analysis of cooperative jamming in cognitive radio  networks with secrecy constraints," \emph{IEEE Wireless Commun. Lett.}, vol. 4, no. 4, pp. 437-440, Aug. 2015.

\bibitem{IEEEhowto:21}
T. Zhang, W. Chen, and F. Yang, ``Balancing delay and energy efficiency in energy harvesting cognitive radio networks: a stochastic stackelberg game approach," \emph{IEEE Trans. Cogn. Commun. Netw.}, vol. 3, no. 2, pp. 201-216, Jun. 2017.

\bibitem{IEEEhowto:22}
J. Denis, M. Pischella, and D. Le Ruyet, ``Energy-efficiency-based resource allocation framework for cognitive radio network with FBMC/OFDM," \emph{IEEE Trans. Veh. Technol.}, vol. 66, no. 6, pp. 4997-5013, Jun. 2017.

\bibitem{IEEEhowto:23}
J. Ouyang, M. Lin, Y. Zou, W.-P. Zhu, and D. Massicotte, ``Secrecy energy efficiency maximization in cognitive radio networks," \emph{IEEE Access.}, vol. 5, pp. 2641-2650, Feb. 2017.

\bibitem{IEEEhowto:24}
D. W. K. Ng, E. S. Lo, and R. Schober, ``Energy-efficient resource allocation for secure OFDMA systems," \emph{IEEE Trans. Veh. Technol.}, vol. 61, no. 6, pp. 2572-2585, Jul. 2012.

\bibitem{IEEEhowto:25}
R. Corvaja and A. G. Armada, ``Phase noise degradation in massive MIMO downlink with zero-forcing and maximum ratio transmission precoding," \emph{IEEE Trans. Veh. Technol.}, vol. 65, no. 10, pp. 8052-8059, Oct. 2016.

\bibitem{IEEEhowto:26}
A. EI-Shafie, D. Niyato, and N. AI-Dhahir, ``Security of rechargeable energy-harvesting transmitters in wireless networks," \emph{IEEE Wireless Commun. Lett.}, vol. 5, no. 4, pp. 384-387, Aug. 2016.

\bibitem{IEEEhowto:27}
A. Khisti and G. W. Wornell, ``Secure transmission with multiple antennas part II: the MIMOME wiretap channel," \emph{IEEE Trans. Inf. Theory,} vol. 56, no. 11. pp. 5515-5532, Nov. 2010.

\bibitem{IEEEhowto:28}
A. Mukherjee, S. Fakoorian, J. Huang, \emph{et al.}, ``Principles of physical layer security in multiuser wireless networks: a survey," \emph{IEEE Commun. Surv. Tut.}, vol. 16, no. 3, pp. 1550-1573, 2014.

\bibitem{IEEEhowto:29}
L. Xu and A. Nallanathan, ``Energy-efficient chance-constrained resource allocation for multicast cognitive OFDM network," \emph{IEEE J. Sel. Areas Commun.}, vol. 34, no. 5, pp. 1298-1306, May 2016.

\bibitem{IEEEhowto:30}
M. El-Halabi, T. Liu, and C. N. Georghiades, ``Secrecy capacity per unit cost," \emph{IEEE J. Sel. Areas Commun.}, vol. 31, no. 9, pp. 1909-1920, Sep. 2013.

\bibitem{IEEEhowto:31}
A. Mukheriee and A. Swindlehurst, ``Detecting passive eavesdroppers in the MIMO wiretap channel," in \emph{Proc. IEEE ICASSP,} Tokyo, Japan, Mar. 2012, pp. 2809-2812.

\bibitem{IEEEhowto:32}
J. Huang and A. L. Swindlehurst, ``Cooperative jamming for secure communications in MIMO relay networks," \emph{IEEE Trans. Signal Process.}, vol.59, no.10, pp. 4871-4884, Oct. 2011.

\bibitem{IEEEhowto:33}
H. Guo, Z. Yang, L. Zhang, J. Zhu, and Y. Zou, ``Power-constrained secrecy rate maximization for joint relay and jammer selection assisted wireless networks," \emph{IEEE Trans. Commun.}, vol. 65, no. 5, pp. 2180-2193, May 2017.

\bibitem{IEEEhowto:34}
H. Gao, M. Wang, and T. Lv, ``Energy efficiecy and spectrum efficiency tradeoff in the D2D-enabled HetNet," \emph{IEEE Trans. Veh. Technol.}, vol. 66, no. 11, pp. 10583-10587, Nov. 2017.

\bibitem{IEEEhowto:35}
H. Niu, D. Guo, Y. Huang, and B. Zhang, ``Robust energy efficiency optimization for secure MIMO SWIPT systems with non-linear EH model," \emph{IEEE Commun. Lett.}, vol. 21, no. 12, pp. 2610-2613, Dec. 2017.

\bibitem{IEEEhowto:36}
J. Ouyang, M. Lin, W. P. Zhu, T. Hong, and B. Xu, ``Distributed-relay beamforming for secrecy energy efficiency with coordinated eavesdroppers," \emph{IEEE Commun. Lett.}, no. 99, pp. 1-4, 2018.

\bibitem{IEEEhowto:37}
H. H. Kha, H. D. Tuan, and H. H. Nguyen, ``Fast global optimal power allocation in wireless networks by local D.C. programming," \emph{IEEE Trans. Wireless Commun.}, vol. 11, no. 2, pp. 510-515, Feb. 2012.

\bibitem{IEEEhowto:38}
H. Wang, J. Wang, G. Ding and Z. Han, ``D2D communications underlaying wireless powered communication networks," \emph{IEEE Trans. Veh. Technol.}, vol. 67, no. 8, pp. 7872-7876, Aug. 2018.

\bibitem{IEEEhowto:39}
C. Kai, H. Li, L. Xu, Y. Li and T. Jiang, ``Energy-efficient device-to-device communications for green smart cities," \emph{IEEE Trans. Ind. Informat.}, vol. 14, no. 4, pp. 1542-1551, Apr. 2018.

\bibitem{IEEEhowto:40}
M. Grant and S. Boyd, ``CVX: Matlab software for disciplined convex programming, version 2.1," [Online:] http://cvxr.com/cvx, Mar. 2014.

\bibitem{IEEEhowto:41}
W. Dinkelbach, ``On nonlinear fractional programming," \emph{Manage. Sci.}, vol. 13, no. 7, pp. 492-498, Mar. 1967.

\bibitem{IEEEhowto:42}
Z.-Q. Luo, W.-K. Ma, A. M.-C. So, Y. Ye, and S. Zhang, ``Semidefinite relaxation of quadratic optimization problems," \emph{IEEE Signal Process. Mag.}, vol. 27, no. 3, pp. 20-34, May 2010.

\bibitem{IEEEhowto:43}
K. An, M. Lin, J. Ouyang, and W.-P. Zhu, ``Secure transmission in cognitive satellite terrestrial networks," \emph{IEEE J. Sel. Areas Commun.}, vol. 34, no. 11, pp. 3025-3037, Nov. 2016.

\bibitem{IEEEhowto:44}
H. A. Suraweera, P. J. Smith, and M. Shafi, ``Capacity limits and
performance analysis of cognitive radio with imperfect channel knowledge," \emph{IEEE Trans. Veh. Technol.}, vol. 59, no. 4, pp. 1811-1822, May 2010.



\end{thebibliography}
\end{document}